\documentclass[usenatbib]{mnras}

\pdfoutput=1
\usepackage{graphicx}
\usepackage{caption}
\usepackage{latexsym}
\usepackage{amsfonts,amsmath,amssymb}
\usepackage{url}
\usepackage[utf8]{inputenc}
\usepackage{fancyref}
\usepackage{hyperref}
\hypersetup{colorlinks=true,pdfborder={0 0 0}, citecolor= blue}
\usepackage{natbib}
\usepackage{multirow}
\usepackage{pdflscape}

\newcommand{\civ}{C~{\textsc IV}}  
\newcommand{\mgii}{Mg~{\sc II}}
\newcommand{\ciii}{C~{\sc III]}}
\newcommand{\heii}{He~{\sc II}}
\newcommand{\oiii}{O~{\sc III]}}
\newcommand{\aliii}{Al~{\sc III}}
\newcommand{\siiii}{Si~{\sc III]}}
\newcommand{\lya}{Ly$\alpha$}
\newcommand{\siiv}{Si~{\sc IV}}

\newcommand{\kms}{${\rm km~s^{-1}}$}

\begin{document}
\title{Insights into Quasar UV Spectra Using Unsupervised Clustering Analysis}

\author[A. Tammour et al. ]{A. Tammour\thanks{atammour@uwo.ca}$^1$, 
S. C. Gallagher$^{1,2}$, 
M. Daley$^3$, \& 
G. T. Richards$^4$ \\ 
$^1$ Department of Physics and Astronomy, University of Western Ontario, 1151 Richmond St, London, ON, N6A 3K7, Canada \\ 
$^2$ Centre for Planetary and Space Exploration, University of Western Ontario,1151 Richmond St, London, ON, N6A 3K7, Canada \\
$^3$ Department of Computer Science, University of Western Ontario,1151 Richmond St, London, ON, N6A 3K7, Canada \\
$^4$ Department of Physics, Drexel University, Philadelphia, PA 19104, USA}
\maketitle

\begin{abstract}
Machine learning techniques can provide powerful tools to detect patterns in multidimensional parameter space. 
We use K-means --a simple yet powerful unsupervised clustering algorithm which picks out structure in unlabeled data-- to study a sample of quasar UV spectra from the Quasar Catalog of the 10$^{\rm th}$ Data Release of the Sloan Digital Sky Survey (SDSS-DR10) of \citet{paris14}.
Detecting patterns in large datasets helps us gain insights into the physical conditions and processes giving rise to the observed properties of quasars. 
We use K-means to find clusters in the parameter space of the equivalent width (EW), the blue- and red-half-width at half-maximum (HWHM) of the \mgii~2800 \AA\ line, the \civ~1549 \AA\ line, and the \ciii~1908\AA\ blend in samples of Broad Absorption-Line (BAL) and non-BAL quasars at redshift 1.6--2.1.
Using this method, we successfully recover correlations well-known in the UV regime such as the anti-correlation between the EW and blueshift of the \civ\ emission line and the shape of the ionizing Spectra Energy distribution (SED) probed by the strength of \heii\ and the \siiii/\ciii\ ratio.
We find this to be particularly evident when the properties of \ciii\ are used to find the clusters, while those of \mgii\ proved to be less strongly correlated with the properties of the other lines in the spectra such as the width of \civ\ or the \siiii/\ciii\ ratio.
We conclude that unsupervised clustering methods (such as K-means) are powerful methods for finding ``natural'' binning boundaries in multidimensional datasets and discuss caveats and future work.
\end{abstract}


\section{Introduction}
\label{sec:intor}

Among the different types of Active Galactic Nuclei (AGN), quasars stand out as the most luminous with typical bolometric luminosities $> 10^{46} {\rm erg~ s}^{-1}$.
Accretion onto a supermassive black hole is now accepted as the main mechanism that powers quasars \citep[][]{shakura73}.
Despite having remarkably similar spectra that typically show a distinct strong blue continuum and broad emission lines with velocity widths of 1000s \kms, quasar spectra exhibit subtle but strong trends among their emission lines and continua that persist over a wide range of wavelengths and luminosities.
These repeated patterns allow us to probe the complex physical processes taking place in quasars' inner regions. 

For example, the profiles of some of the broad high ionization lines (e.g., \civ~$\lambda 1550$) exhibit some structure which indicates that in addition to the Doppler broadening of the line by the central black hole's gravitational field, a non-virial component of motion in the gas (inflow or outflow) is at work creating this structure seen as blue asymmetries or blueshifts of the peak from the systemic redshift measured in lower ionization lines \citep[such as \mgii, e.g.,][]{richards02}.
Another notable example is the Baldwin Effect \citep[][]{baldwin77} seen as an anti-correlation between the strength of \civ~$\lambda$1550 (measured by its equivalent width, EW) and the continuum luminosity at 1550 \AA.

More recently it has been demonstrated that \civ\ EW is anti-correlated with its blueshift and both of those quantities are tied to the X-ray hardness of the quasar in a sense that objects that are soft in the X-ray (characterized by the spectral index $\alpha_{ox}$\footnote{The spectral index $\alpha_{ox}= 0.384 \log(f_{2 {\rm keV}}/f_{2500})$, measures the slope of the flux densities at 2 keV and rest frame wavelength 2500 \AA. The quantity 0.384 is the logarithm of the ratio of the frequencies at which the flux densities are measured.}) are likely to have weaker \civ\ and larger blueshifts \citep[e.g.,][]{richards02, leighlymoore04, leighly04, richards11}.
Moreover, an intrinsic fraction of $\sim 20\%$ of quasars of optically selected samples display broad ($\Delta V > 2000$ \kms) UV absorption features; broad absorption line (BAL) quasars \citep[e.g.,][]{weymann91, hewett03}.
BAL quasars exhibit relatively weak X-ray emission \citep[e.g.,][]{gallagher06} and show blueshifted broad absorption troughs with large velocity offsets indicating an absorbing medium moving outward with velocities up to 25,000 \kms \citep[e.g.,][]{weymann91}.  
A two-component origin of quasar  broad-lines can successfully account for many of their observed properties such as the blueshift-EW anti-correlation of high-ionization lines (e.g., \civ) and its connection with the SED hardness in which broad-lines are a combination of emission from the accretion disk and emission from a fast outflowing gas accelerated by radiation pressure that emerges close to the accretion disk \citep[e.g.,][]{collin88,murray98,proga00}.
In this scenario, BAL quasars are seen through the angle covered by the winds and their weaker X-ray emission is a consequence of the winds filtering the energetic photons out of the ionizing radiation \citep[e.g.,][]{proga00,leighly04,leighlymoore04}. 

\subsection{Patterns in Multi-dimensional Space}
Looking for systematic patterns appearing repeatedly among the measured variables in large multi-dimensional datasets of quasars can help elucidate the physical driver behind those trends.
It is often the case in spectral studies that finding those patterns requires binning the dataset using one or two important parameters then comparing the properties of objects among the different bins by stacking the spectra to create (median or mean) composite spectra \citep[e.g.,][]{croom02, richards11, hill14, shen14, baskin15,tammour15}.
The binning has been done in most cases in a two-dimensional space with fixed boundaries which follow ``traditional'' cuts such as 2000 or 4000 \kms\ for the width of H$\beta$ \citep[e.g.,][]{sulentic07,tammour15} which is not an unreasonable choice as these parameters and boundaries are found to best constrain the properties of the objects under study and to be separating meaningful classes of AGN \citep[e.g.,][]{boroson92, boroson02, sulentic00}.
However, it may be possible to separate objects in a more optimal way when it comes to identifying physical differences.

One notable method used to study multidimensional parameter space in quasar samples is Principal Component Analysis --a statistical method that takes a multidimensional dataset and finds the orthogonal axes (i.e., Eigenvectors) that minimize the variance along each projection \citep{elements, astroml}.
This approach was able to uncover interesting correlations among quasar properties \citep[e.g.,][]{boroson92, boroson02, yip04}.
\citet{boroson92}, for example, found a strong inverse correlation between the strength of the Fe II~$\lambda$4570 complex and the strength of [O III]~$\lambda$5007.
This correlation, has been consistently found in other quasar samples and is thought to originate from the Eddington accretion rate $L_{\rm bol}/L_{\rm Edd}$ \citep[e.g.,][]{boroson92, sulentic00, boroson02, shen14}.

In this work , we explore the use of unsupervised clustering analysis to find patterns among quasar spectral properties in the UV regime.
Unlike supervised learning which uses a labelled training set to assign labels to unlabelled data, unsupervised learning does not require previous knowledge of labels --- it is ``learning without a teacher'' \citep[][Ch. 14]{elements}.
Clustering analysis is one of the unsupervised learning techniques that aims to find structure in multidimensional data space \citep{pattern, elements}. 
In unsupervised clustering, the data is assigned to clusters according to a chosen metric of similarity such as the Euclidian distance (e.g, the K-means algorithm) or non-Euclidian metrics (e.g., Density Based Spatial Clustering, DBSC) \cite[e.g.,][]{elements, astroml}.
Despite its robustness to reveal important properties about quasars, PCA assumes orthogonality of the axes defined by the eigenvectors that are a linear combination of the input parameters.
This makes interpreting the eigenvectors often physically rather non-intuitive. 
Clustering of sources by K-means, on the other hand, more closely follows the divisions inherent to the data that we might hope are ultimately due to physics (after accounting for any selection effects).

In \S \ref{sec:selection} we describe our samples and in \S \ref{sec:kmeans} we discuss in detail the K-means algorithm and how we apply it to our quasar samples.
\S \ref{sec:results} includes our results and in \S \ref{sec:caveats} we discuss some of the caveats of this technique. 
A summary and conclusion are given is \S \ref{sec:conclusion}.

Throughout this work we use: H$_0$ = 70 km s$^{-1}$, $\Omega_M$ = 0.3, $\Omega_{\Lambda}$ = 0.7 \citep[][]{spergel03}.

\section{Analysis}

\subsection{Sample Selection}
\label{sec:selection}

Our starting point in selecting this sample is the \citet{paris14} Quasar Catalog of the 10$^{th}$ Data Release of the Sloan Digital Sky Survey (SDSS-DR10).
The large size of this catalog and its improved measurements from those of the SDSS automated pipeline make it ideal for our statistical study of quasar UV spectra.  
The catalog contains 166,583 quasars with measurements of emission and absorption lines such as the amplitude (AMP), equivalent width (EW), full-width at half-maximum (FWHM) and the red and blue half-width at half-maxima (RHWHM and BHWHM) of \mgii, the \ciii\ blend and \civ.
These measurements were obtained with a linear combination of a set of principal components (eigenspectra) to fit the the full spectrum in a similar fashion to the DR9 Quasar Catalog \citep{paris12}.
We choose to restrict the selection to objects within the redshift range of 1.6 to 2.2 (using the PCA redshift values in the \citet{paris14} estimated from fitting for the whole spectrum) to allow us to study some of the interesting UV features such as: \lya~$\lambda$1215,  \siiv~$\lambda$1396, \civ~$\lambda1550$, the \heii~$\lambda1640$ and \oiii~$\lambda1664$ blend, the \aliii~$\lambda1857$, \siiii~$\lambda1892$ and \ciii~$\lambda1908$ blend, and \mgii~$\lambda2800$ \citep{vandenberk01}.
We note that the redshift values listed in the catalogue could potentially contain inaccurate estimates.
This is mainly because high and low ionization lines are commonly seen to be shifted (from systemic velocities) in different proportions. 
However, in case the redshifts values are systematically wrong, this will lead to biased measurements that will reveal themselves as separate clusters that can potentially lead to new insights into the physical properties of these objects as we discuss in \S \ref{sec:results}.

We select the initial sample (main sample, \S \ref{sec:mg2}, \ref{sec:c4}, and \ref{sec:c3}) based on the following criteria: EW is $> 0$ for each of the lines \civ, \ciii\ and \mgii, the EW error is $< 10\%$ of the measured EW for all of the three lines, the BAL flag from visual inspection (BAL\_FLAG\_VI) is set to 0 (no BAL quasars), and finally we use a signal-to-noise minimum of  S/N $> 3$ at 1700 \AA.

\begin{figure}
\includegraphics[width=\linewidth]{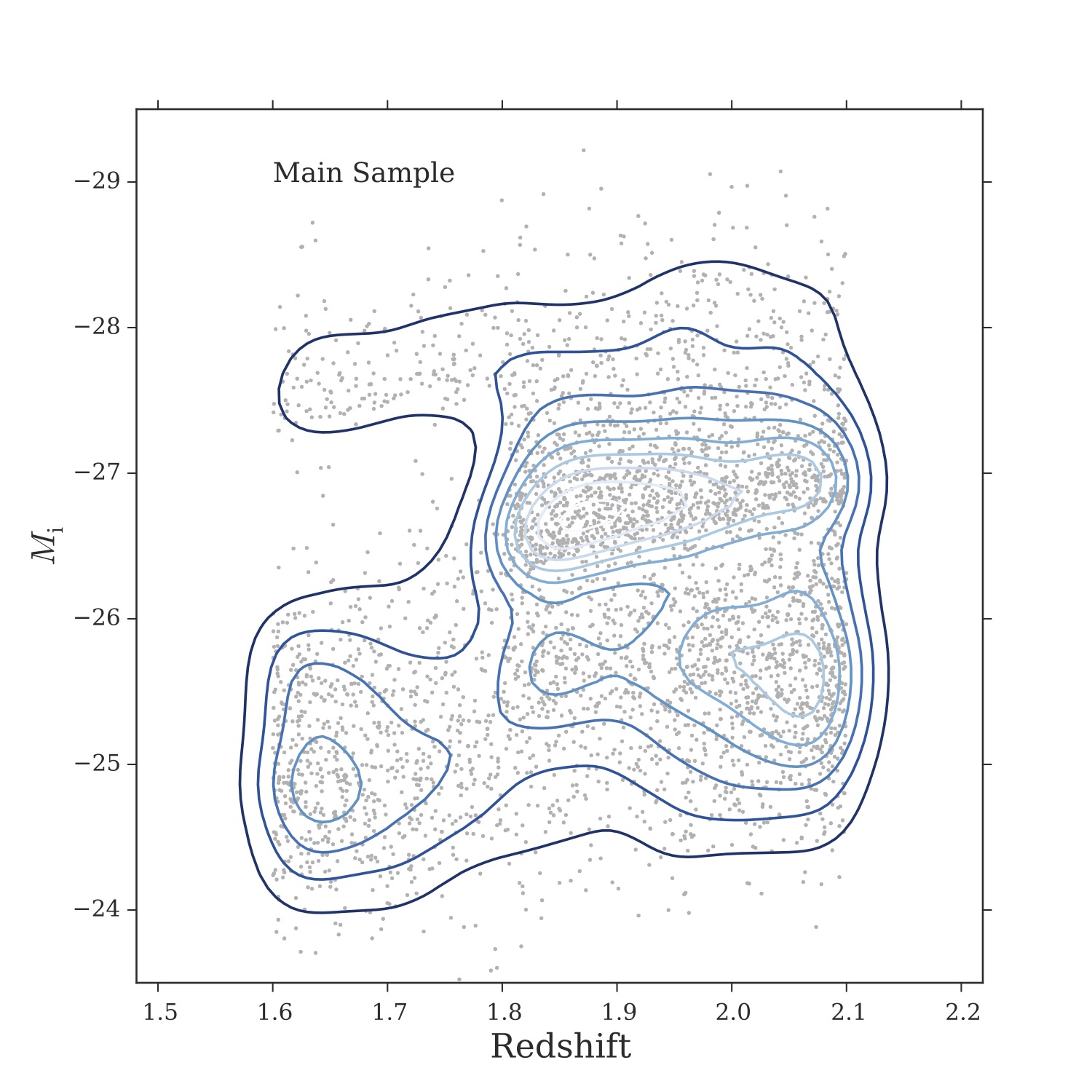}
\caption{Distribution of $M_{\rm i}$ vs. $z$ for the objects in our sample.
The objects in our main sample are shown with grey dots and blue contours. 
Given the narrow range of redshift and the SDSS selection, the luminosity distribution is relatively flat with $z$.}
\label{fig:mi_z}
\end{figure}

This selection gives us a sample of 4110. 
Fig. \ref{fig:mi_z} shows the distribution of M$_i$ vs. redshift in the main sample.
We visually examine the individual spectra of these objects and remove 7 objects with bad spectra (missing flux) and one object that looked like a misidentified BAL quasar. 
In addition to this main sample, we select two other samples: mixed sample with the EW $>0$ and the BAL flag condition relaxed (contains 6463 objects, see \S \ref{sec:mixed}), and BALQ sample with the BAL quasars \emph{only} (contains 1533 objects, see \S \ref{sec:bal}).

We finally note that there were many cases of heavy narrow absorption in \civ\ and sometimes in \mgii\ that were not flagged as BAL quasars in the catalog.
We decide to keep those objects in our samples because we wanted to test if the algorithm is able to isolate those objects from the larger quasar population; see \S \ref{sec:results} for more discussion about this point.

\subsection{Unsupervised Clustering with the K-means Algorithm}
\label{sec:kmeans}

K-means is one of the most widely used unsupervised clustering algorithms \citep{elements, pattern}.
Part of the popularity of this algorithm is due to its simple and intuitive design.
K-means aims to minimize the ``inertia criteria'' (i.e., within-cluster sum of squares) given by:
\begin{equation}
J= \sum_{n=1}^{N} \sum_{k=1}^{K} \mbox{min} (||x_n- \mu_k||^2),
\label{eqn:sos}
\end{equation}
where $\mu$ is the mean in cluster $k$, $N$ is the total number of data points (samples) and $K$ is the number of clusters.
In a nutshell, the algorithm starts by assigning a fixed number of random points to serve as centroids for the clusters to be found.
It then proceeds to assign data points that are closer to each centroid to a cluster.
Those new cluster members are then used to calculate a new mean which becomes the new cluster centroid.
The distances between the new centroids and the cluster members are calculated again and the members are reassigned to the clusters with the smallest distance to their centroids.
The algorithm continues to iterate this calculating and reassigning process until it converges.
A good visual illustration of this process is given in \citet[their Fig. 9.1;][]{pattern}.

To perform the K-means clustering, we use {\tt scikit-learn}\footnote{Using the sklearn.cluster.KMeans package: \url{http://scikit-learn.org/stable/modules/generated/sklearn.cluster.KMeans.html\#sklearn.cluster.KMeans}} \citep[][]{scikit-learn} applied to the EW, BHWHM, and RHWHM measurements provided by the \citet{paris14} catalogue for each of the \civ, \ciii, and \mgii\ lines separately.
After extensive experimentation, we chose to focus on these three parameters (features) because they adequately describe the lines in terms of strength and structure (asymmetry).
In the case of the \ciii\ blend, the blue- and red-HWHM also serve as a measure of the strengths of the \siiii\ and \aliii\ lines blue-ward of and often heavily blended with the \ciii\ line (we discuss this further in \S \ref{sec:c3}).

The only parameter that this algorithm requires beforehand is $K$: the number of clusters that we want (expect) our data to be grouped into.
Determining this $K$ is not necessarily a clear-cut exercise as the ``ground truth'' in our case is not known, i.e., the data points are not labeled.
However, heuristic approaches coupled with a good understanding of the dataset and its features are sufficient to determine a range of optimal values of $K$.
We fully expect that the clustering parameters have continuous distributions, perhaps with breaks due more to selection effects than anything.  
What the clustering does here, is to provide a quantifiable and less arbitrary way of breaking those objects up into groups that can be used to follow the trends in that continuum.

\begin{figure}
\centering
\includegraphics[width= \linewidth]{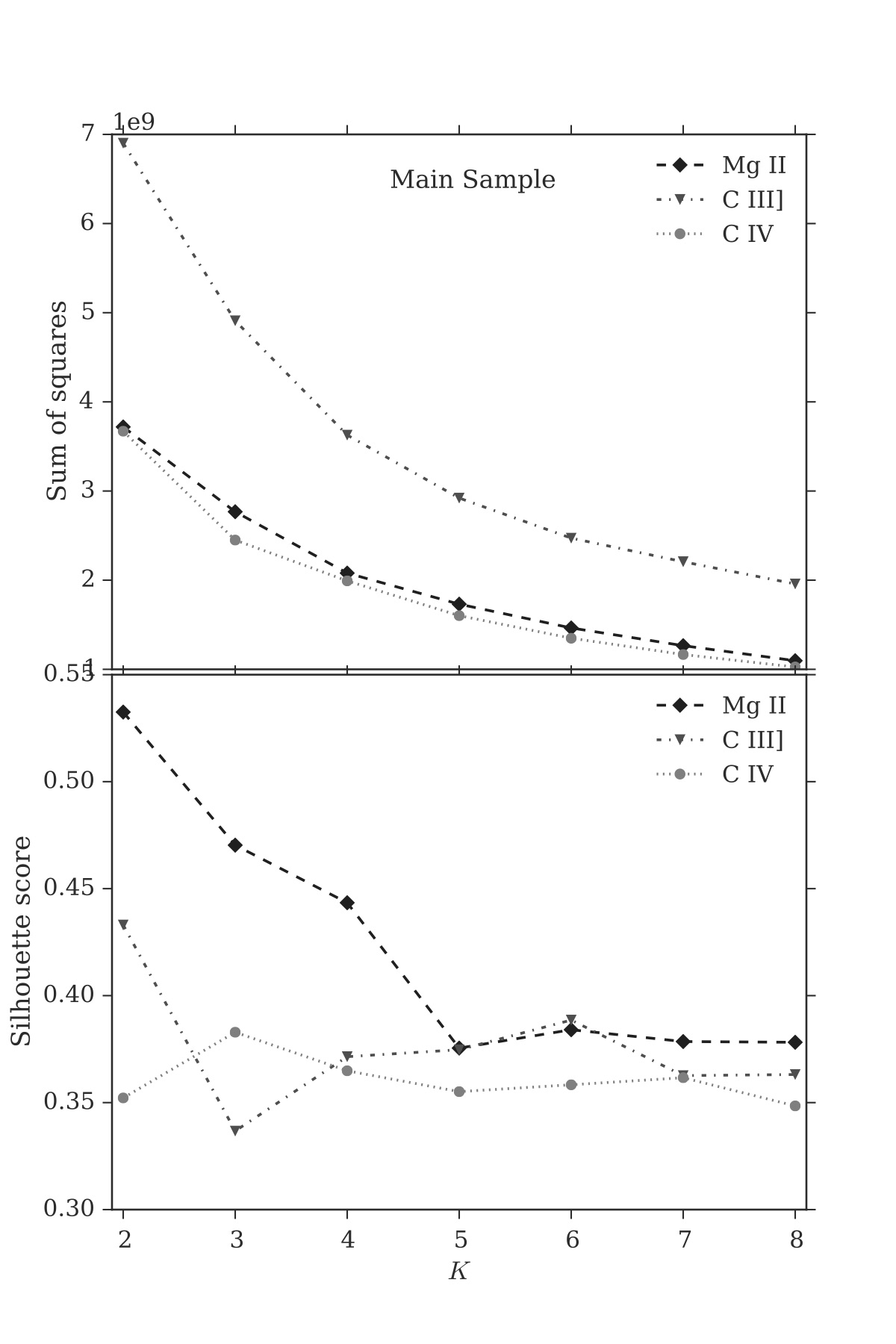}
\caption{Top: The sum of the squared distances between data points and their cluster centroid (eqn. \ref{eqn:sos}) calculated for $K=$ 2 to 8 and using EW, RHWHM, and BHWHM for each of the \ciii, \civ\ and \mgii\ line (blend) separately.
For our samples, a range of 3 to 6 is sufficient to capture the curve minimum.
Bottom: Average Silhouette scores (eqn. \ref{eqn:silhouette}) for all data points calculated for the clustering done on each of the three emission lines separately using $K=$ 2--8. 
Higher scores indicate less overlap among clusters. 
For \mgii, $K = $2, 3 and 4 give the best separation.
While \ciii\ requires $K = 2$ or $K > 4$ and for \civ\ $K > 3$ gives the best separated clusters.}
\label{fig:scores_main}
\end{figure}

In Fig. \ref{fig:scores_main}, we show two metrics that describe the ``goodness of fit'' for different values of $K$ starting from $K=2$ to $K=8$.
The top panel gives the sum of the square distances (errors) between data points (samples) and their cluster centroid for each attempt.
$K$ at the ``elbow'' of Fig. \ref{fig:scores_main} (top panel) is the one that minimizes the inertia given in equation \ref{eqn:sos}. 
Ideally, one can get a good estimate of $K$ by looking for the point where the curve bends (forms an elbow), but in most ``real'' datasets the curve is smooth and one can only find a range of possible $K$ values that minimizes the square error.
The lower panel gives the Silhouette Score, a metric that measures the isolation of clusters by comparing the mean of the distances between a sample and its fellow cluster members, $a$, and the mean of distances between this sample and the ones in the nearest neighbour cluster, $b$ \citep{scikit-learn, data_mining}:
 \begin{equation}
 s= \frac{b-a}{max(a,b)} .
 \label{eqn:silhouette}
 \end{equation}
This score, $s$, ranges from 1 for a dense, well-defined cluster to -1 for a less dense cluster that overlaps with the ones surrounding it.

Fig. \ref{fig:scores_main} suggests that 3 or 4 clusters are reasonable for \civ\ and \mgii\ while more clusters might be needed for \ciii.
We proceed to use $K=$ 3 to 6 for all 3 lines and we discuss in \S \ref{sec:results} cases where adding an extra cluster sometimes serves to isolate outliers and create a ``cleaner'' cluster.
Table \ref{tbl:num_clstrs} shows a breakdown of the clustering results for each emission line or blend for $K=$ 3, 4, 5 and 6.

\clearpage
\begin{landscape}
\begin{table}
\centering
\small
\caption{Number of objects in each cluster formed with K-means applied in the main sample to measurements of EW, BHWHM, and RHWHM of \ciii, \civ, and \mgii\ separately using $K=$ 3, 4, 5 and 6. 
For each cluster, we also give the coordinates of the cluster centroids in the (EW,BHWHM,RHWHM) space. 
The labels ``a'', ``b'', ``c'', etc. refer to the same clusters and composites shown for the rest of this paper. 
Clusters are ordered according to the increasing values of BHWHM of their centroids.
Clusters/composites are labeled with the name of the line used for the clustering and the number of clusters used. 
For example, \ciii-5d is cluster d from the clustering done using $K=5$ in the \ciii\ parameter space.
In Appendix \ref{sec:rep_plots} we show that the algorithm is able to find the same unique centroids for the clusters after up to 50 repeats of running K-means.}
\label{tbl:num_clstrs}
	\begin{tabular}{lccccccc}
	\hline
	\hline
	& &                            \multicolumn{6}{|c|}{Number of objects} \\
	&  &                           \multicolumn{6}{|c|}{EW (\AA), BHWHM (km s$^{-1}$), RHWHM (km s$^{-1}$)} \\
	\hline
	Line & $K$ &a & b & c & d & e &f \\
	\hline
		\multirow{8}{*}{\ciii}	&\multirow{2}{*}{3}& 1437&2309&356& -- & -- & --\\ 
					& & (24.35,1783.97,2534.25) & (26.85,2754.39,3821.80) & (29.68,3376.13,6692.33) &  -- & -- & --\\ 
					
					& \multirow{2}{*}{4}& 980&1919&328&875&  -- & --\\
					&   &  (23.76,1711.87,2226.60) & (26.75,2177.15,3857.84) & (29.66,3230.03,6828.22) & (26.50,3668.96,3455.78) &   -- & --\\
					
					& \multirow{2}{*}{5}& 392&1556&1059&288&807&  --\\
					&   & (22.90,1489.56,1471.68) & (25.44,1988.50,3133.32) & (27.47,2365.86,4354.74) & (29.67,3268.35,7016.40) & (26.46,3718.25,3425.02) &   --\\
					
					& \multirow{2}{*}{6}& 375&1468&1063&209&774&213  \\
					&   & (22.91,1473.23,1437.42) & (25.40,1937.23,3126.74) & (27.52,2345.50,4337.00) & (29.65,2770.46,7281.64) & (25.91,3460.89,3191.11) & (28.90,4591.70,5247.12) \\
	\hline
	\multirow{8}{*}{\civ}	& \multirow{2}{*}{3}& 1713&1313&1076& -- & --& --\\
					&    & (43.39,1746.90,1958.73) & (31.04,2044.52,3507.01) & (38.35,3304.51,2698.20) &  -- & --& --\\
					
					& \multirow{2}{*}{4}& 1284&1270&1020&528& -- & --\\ 
					&   & (43.97,1562.68,1880.82) & (30.62,1971.40,3453.93) & (41.03,2775.14,2248.06) & (36.29,3582.87,3349.97) & -- & --\\ 
					
					& \multirow{2}{*}{5} & 977&1280&486&791&568& -- \\
					&   & (45.94,1527.30,1688.75) & (33.91,1945.03,2872.95) & (29.37,1999.76,4103.41) & (42.30,2892.92,2122.82) & (35.79,3504.51,3279.19) &  -- \\
					
					& \multirow{2}{*}{6}& 856&1060&417&756&730&283 \\
					&  & (46.92,1507.34,1617.35) & (34.27,1769.15,2746.57) & (28.99,1923.20,4188.69) & (33.70,2683.73,3194.66) & (43.22,2859.02,2060.24) & (37.99,4060.17,3185.96) \\
	\hline
	\multirow{8}{*}{\mgii}	& \multirow{2}{*}{3}& 2503&1258&341& -- & -- &-- \\
					&   & (38.14,1795.43,1739.41) & (46.39,3129.40,2591.14) & (51.15,3946.39,4451.96) & -- & -- &-- \\
					
					& \multirow{2}{*}{4}&2176&1271&241&414& -- & --\\
					&   & (37.42,1757.32,1630.73) & (44.83,2585.56,2628.85) & (51.47,3629.85,4850.80) & (49.41,4327.26,2591.17) &  -- & --\\
					
					& \multirow{2}{*}{5}&1468&1371&707&159&397& -- \\
					&   & (35.36,1589.28,1498.43) & (42.53,2262.72,2087.34) & (46.65,2818.55,3085.96) & (53.23,3799.77,5284.70) & (49.38,4393.77,2639.92) & -- \\
					
					&\multirow{2}{*}{6} & 1482&1366&622&77&366&189 \\
					&   & (35.42,1596.65,1496.80) & (42.54,2235.09,2131.36) & (46.98,2830.91,3192.47) & (51.61,3100.29,6065.12) & (48.02,3989.87,2222.78) & (52.35,4747.26,3891.82) \\
	\hline
	\end{tabular}
\end{table}
\end{landscape}

\subsection{Median Composite Spectra}
\label{sec:compos}
To better visualize the results of the clustering and to examine the properties of the other lines which were not used in the clustering analysis, we create median composite spectra from the objects in each of the clusters we find.
We start by correcting all individual spectra for Galactic extinction and redshift using the extinction at the {\rm g}-band and the redshift values estimated from PCA as quoted in the \citet{paris14} catalogue.
We normalize the individual spectra using the median of the flux between 2360 \AA\ and 2390 \AA\ and apply $3\sigma$ clipping\footnote{Each median spectrum only includes points contributing to the spectrum that are inside the range of $3\sigma$ standard deviations in each wavelength bin.} before they are median-combined.

\section{Results and Discussion}
\label{sec:results}

\subsection{\mgii\ Clusters}
\label{sec:mg2}

\begin{figure*}
  \includegraphics[width=\linewidth]{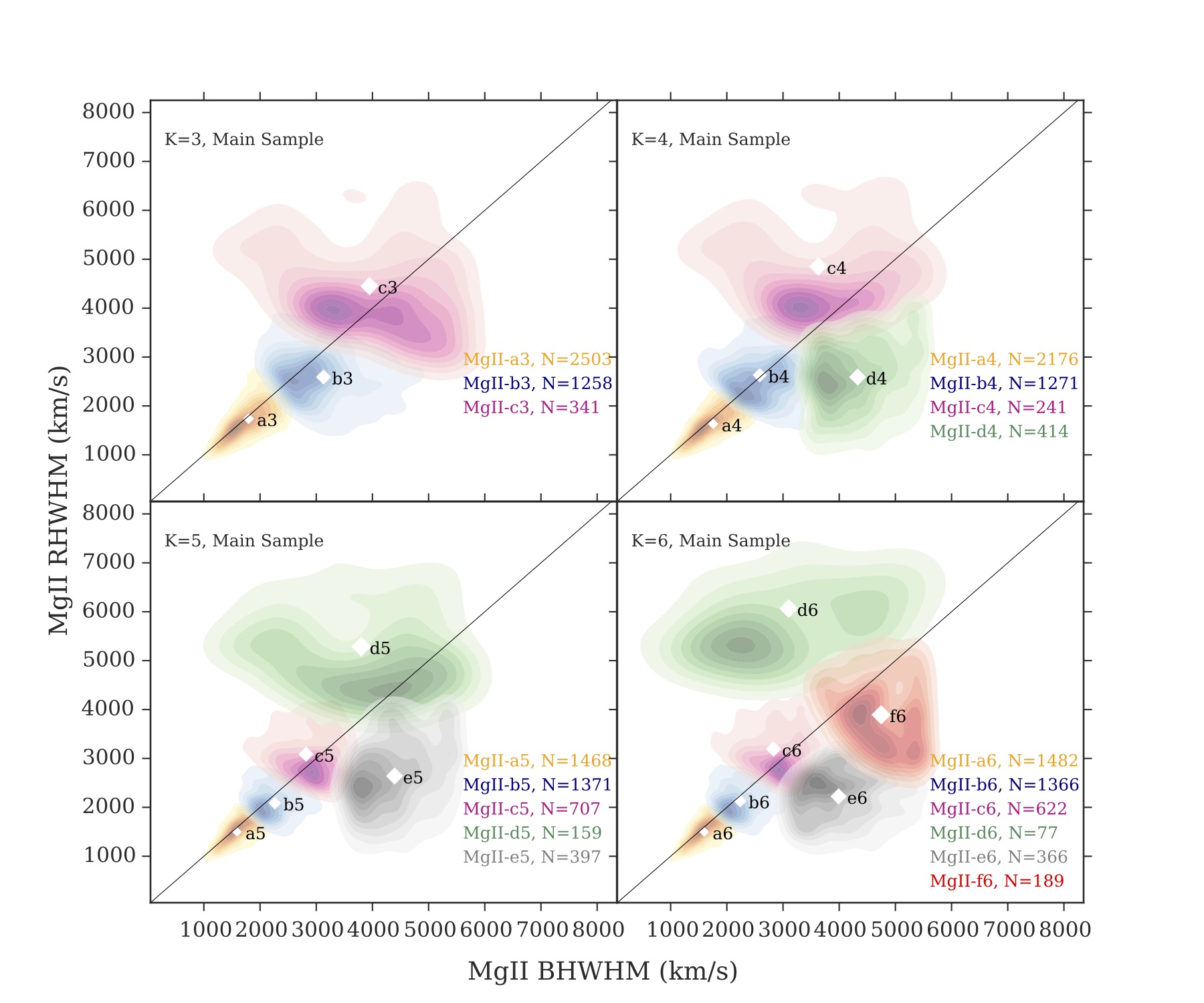}
  \caption{Red- vs. Blue-HWHM plane for clusters found using $K=$ 3 to 6 (as labeled in the top-left corner of each panel) and the EW, BHWHM and RHWHM measurements of \mgii. 
  The cluster labels match the ones given in Table \ref{tbl:num_clstrs}. 
  The white diamonds show the projected locations of the cluster centroids in this plane and their sizes are proportional to the mean EW of each cluster. 
  The diagonal black lines mark the 1 to 1 ratio (symmetric line).}
  \label{fig:mg2_clstrs}
  \end{figure*}

\begin{figure*}
  \includegraphics[width= \linewidth]{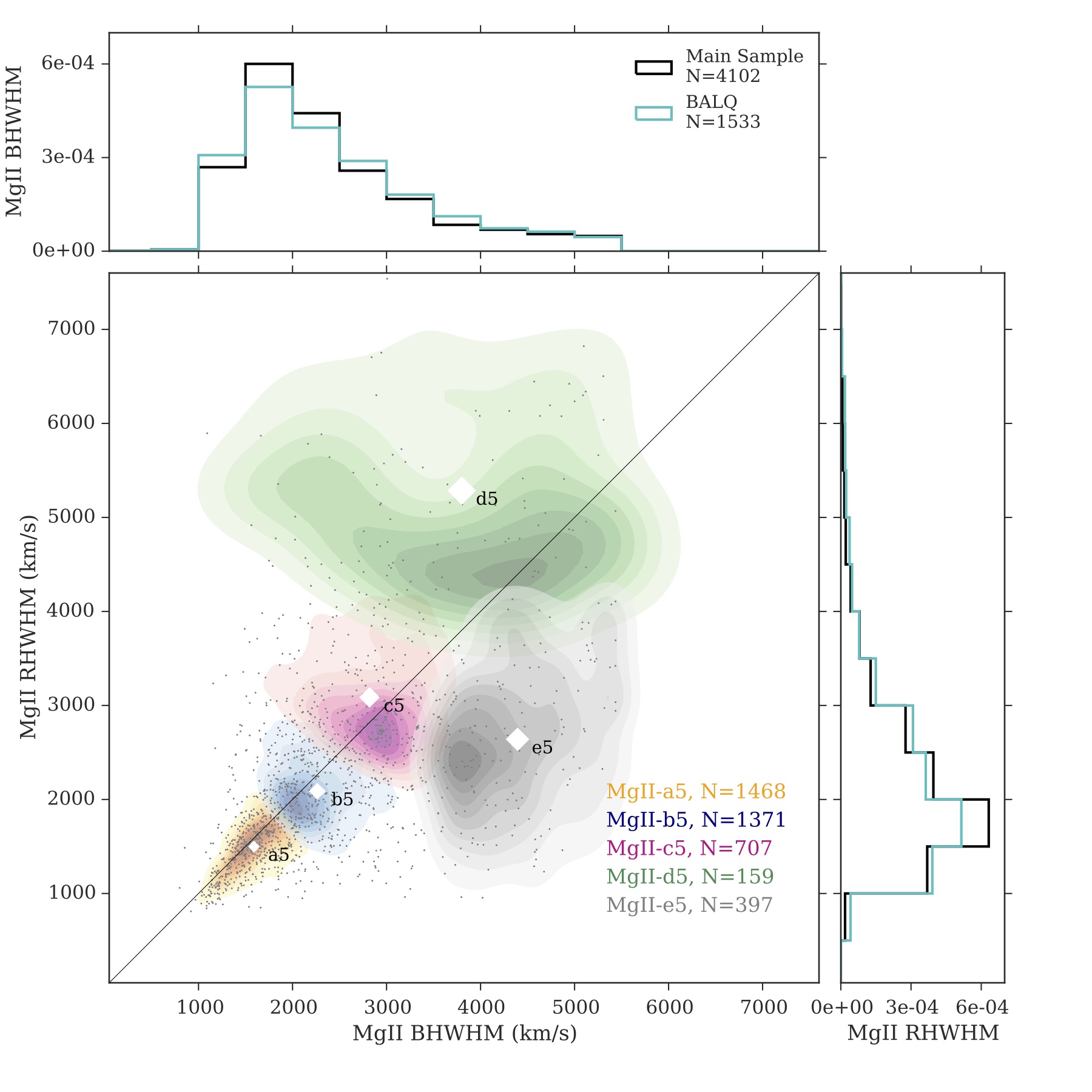}
  \caption{Red- vs. Blue-HWHM of \mgii\ for the $K=5$ clustering run. 
 We over-plot the BAL quasars (not included in the main sample) as grey dots and show the fractional distributions of the \mgii\ blue- and red-HWHM of the main and BALQ samples in the upper and left panels respectively.
It is clear that \mgii\ has a fairly symmetric profile except for the d5 and e5 clusters which we discuss further in the text. 
Both the blue- and red-HWHM of \mgii\ of the BAL quasars have distributions similar to those in the main sample.
The white diamonds show the projected locations of the cluster centroids and their sizes correspond to the mean EW for each cluster.
The average EW of \mgii\ increases with its width as the sizes of the white diamonds appear to increase diagonally with the RHWHM and BHWHM.}
  \label{fig:mg2_k5_kde_bal_hist}
  \end{figure*}

The \mgii\ clusters generated using EW, BHWHM and RHWHM of \mgii\ are shown in Fig. \ref{fig:mg2_clstrs} for $K$ = 3, 4, 5, and 6. 
The figure shows the 2-D plane of the blue- and red-HWHM as the $x$ and $y$-axes.
The figure also shows the mean EWs for each cluster.
While the EW is still contributing to the clustering (see values of the cluster centroid coordinates in Table. \ref{tbl:num_clstrs}), the two width parameters have more contribution in determining the clustering results.
In Fig. \ref{fig:mg2_k5_kde_bal_hist}, we show the $K=5$ run only along with the BALQ sample quasars (not used in the clustering analysis) over-plotted and the distributions of both the full main and BALQ samples.
The \mgii\ line is largely symmetric (with the exception of \mgii-e5 which we discuss below) and its average EW increases in the clusters with its width as Fig. \ref{fig:mg2_k5_kde_bal_hist} shows
The red- and blue-HWHM of \mgii\ appear to have similar distributions in both the BAL and non-BAL quasar samples.

\begin{figure*}
\includegraphics[width=\linewidth]{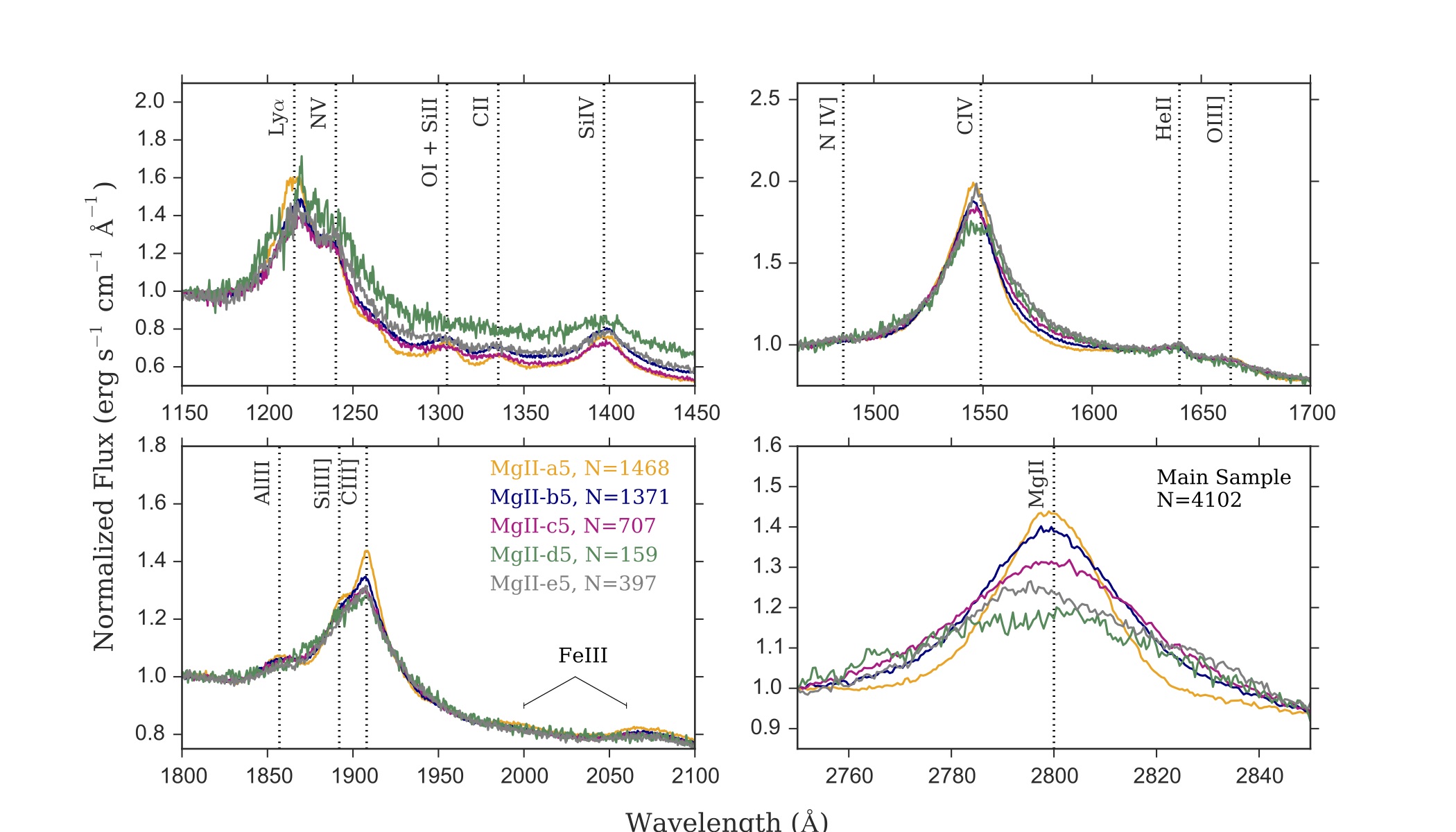}
\caption{Median composite spectra made from the objects in the \mgii\ clusters with $K=5$ shown in Fig. \ref{fig:mg2_k5_kde_bal_hist}. 
The profiles have been normalized locally at the starting wavelength of each panel. 
The numbers in the lower-left panel refer to the number of objects in each composite/cluster (see also Table \ref{tbl:num_clstrs}).
The width of \mgii\ appears to be the main driver for the clustering and is potentially probing average black hole masses in each cluster.
Composites \mgii-d5 and \mgii-e5 have objects with relatively high red- and blue-HWHM respectively as shown in Fig. \ref{fig:mg2_k5_kde_bal_hist}. 
The weak dip in the \civ\ profile in composite \mgii-d5 betrays the presence of narrow absorption features in many objects in this cluster.}
	\label{fig:mg2_k5_profiles}
\end{figure*}

We median-stack objects in each one of the clusters shown in Fig. \ref{fig:mg2_clstrs} and show the results in Fig. \ref{fig:mg2_k5_profiles}.
The projected width of \mgii\ is the predominant quantity driving differences among clusters (as the \mgii\ profiles indicate) and is potentially probing differences in the black hole masses of objects in each cluster.
However, with this change in the widths of \mgii, we do not see any clear corresponding changes in the strengths, widths or asymmetries of any of the other lines in the composites. 
\civ\ for example appears to have nearly identical profiles in all the composites regardless of the width of \mgii.
Because of the small range of luminosity in the sample, the \mgii\ clusters are likely reflecting BH masses.
This lack of correlation between the line widths of \mgii\ and \civ\ is consistent with the known discrepancy in the estimates of black hole masses frequently done using single epoch \civ\ measurements \citep[e.g.,][]{baskin05, shen12}.

The absorbed peak of \civ\ in composite \mgii-d5 (Fig. \ref{fig:mg2_clstrs}) appears to be a result of many narrow absorption lines in \civ\ in this cluster that are evident in the median composite due to the low number of objects (159 objects).
We visually examine the spectra of individual objects in this cluster and find that most of them have very broad \mgii\ profiles and frequent narrow absorption in \mgii\ and/or \civ.
 Most objects in cluster \mgii-e5 in Fig. \ref{fig:mg2_k5_kde_bal_hist} fall below the diagonal causing the skewed blue-ward \mgii\ profile in Fig. \ref{fig:mg2_k5_profiles}.
Visual examination of objects in this cluster showed a few cases where the redshift might have been poorly estimated potentially due to narrow absorption in \civ\ and/or \mgii\ which likely causes the automated fitting to misidentify the line peak. 
Most objects, however, appear to have \civ\ and \ciii\ peaking at systemic redshift. 
We check for discrepancies in the redshift estimates by comparing the PCA redshift (which we used in shifting our spectra to restframe) to those estimated from \mgii\ alone, as given in the \citet{paris14} catalog, and found that they are in good agreement (Table \ref{tbl:redshifts}).
Indeed \mgii\ is known to be one of the most reliable redshift indicator among broad permitted lines in quasar spectra \citep[e.g.,][]{hewett10}.
Our visual inspection also identified an enhancement red-ward of \mgii\ in many of the objects in the \mgii-e5 cluster. 
Such enhancement can be a result of stronger FeII features blended with \mgii\ which could cause the observed blue-skewness of the \mgii\ line.

\begin{table*}
\caption{Comparison between the redshifts estimated from PCA and individual lines as given in the \citet{paris14} catalog for each of the clusters we identify in the \mgii, \civ\ and \ciii\ runs.
We report the statistic and $p$ value of the K-S test which tests the null hypothesis that the two samples are drawn from the same distribution.
Redshift estimates using \mgii\ have similar distributions to those from PCA with high significance in all clusters.
Redshifts from \ciii\ are slightly less identical to those from PCA except for \ciii-e5 which appears to have significantly different redshifts from those estimated from PCA. 
\civ\ redshifts vary in their similarities to those from PCA but are mostly in good agreement with them.
}
\label{tbl:redshifts}
\begin{tabular}{lccccccc}
\hline
\hline
Cluster	&Num Obj	&$\Delta$Z(\mgii)&$p$		&$\Delta$Z(\ciii)	&$p$		&$\Delta$Z(\civ)	&$p$ \\
\hline
\ciii-a5	&392		&0.020	&1.000	&0.018	&1.000	&0.015	&1.000 \\ 
\ciii-b5	&1556	&0.008	&1.000	&0.013	&0.999	&0.019	&0.932 \\
\ciii-c5	&1059	&0.009	&1.000	&0.021	&0.975	&0.029	&0.749 \\
\ciii-d5 	&288		&0.028	&1.000	&0.042	&0.960	&0.059	&0.685 \\
\ciii-e5	&807		&0.020	&0.997	&0.061	&0.098	&0.038	&0.583 \\
\hline
\civ-a5	&977		&0.011	&1.000	&0.012	&1.000	&0.013	&1.000 \\
\civ-b5	&1280	&0.011	&1.000	&0.023	&0.894	&0.026	&0.784 \\
\civ-c5	&486		&0.019	&1.000	&0.035	&0.923	&0.033	&0.952 \\
\civ-d5	&791		&0.011	&1.000	&0.028	&0.916	&0.037	&0.655 \\
\civ-e5	&568		&0.016	&1.000	&0.046	&0.582	&0.048	&0.533 \\ 
\hline
\mgii-a5	&1468	&0.012	&1.000	&0.015	&0.996	&0.026	&0.704 \\
\mgii-b5	&1371	&0.009	&1.000	&0.030	&0.566	&0.028	&0.662 \\
\mgii-c5	&707		&0.018	&1.000	&0.031	&0.879	&0.024	&0.986 \\
\mgii-d5	&159		&0.050	&0.986	&0.038	&1.000	&0.057	&0.956 \\
\mgii-e5	&397		&0.038	&0.935	&0.033	&0.982	&0.028	&0.998 \\
\hline
\end{tabular}
\end{table*}

\subsection{\civ\ Clusters}
\label{sec:c4} 

\begin{figure*}
  \includegraphics[width= \linewidth]{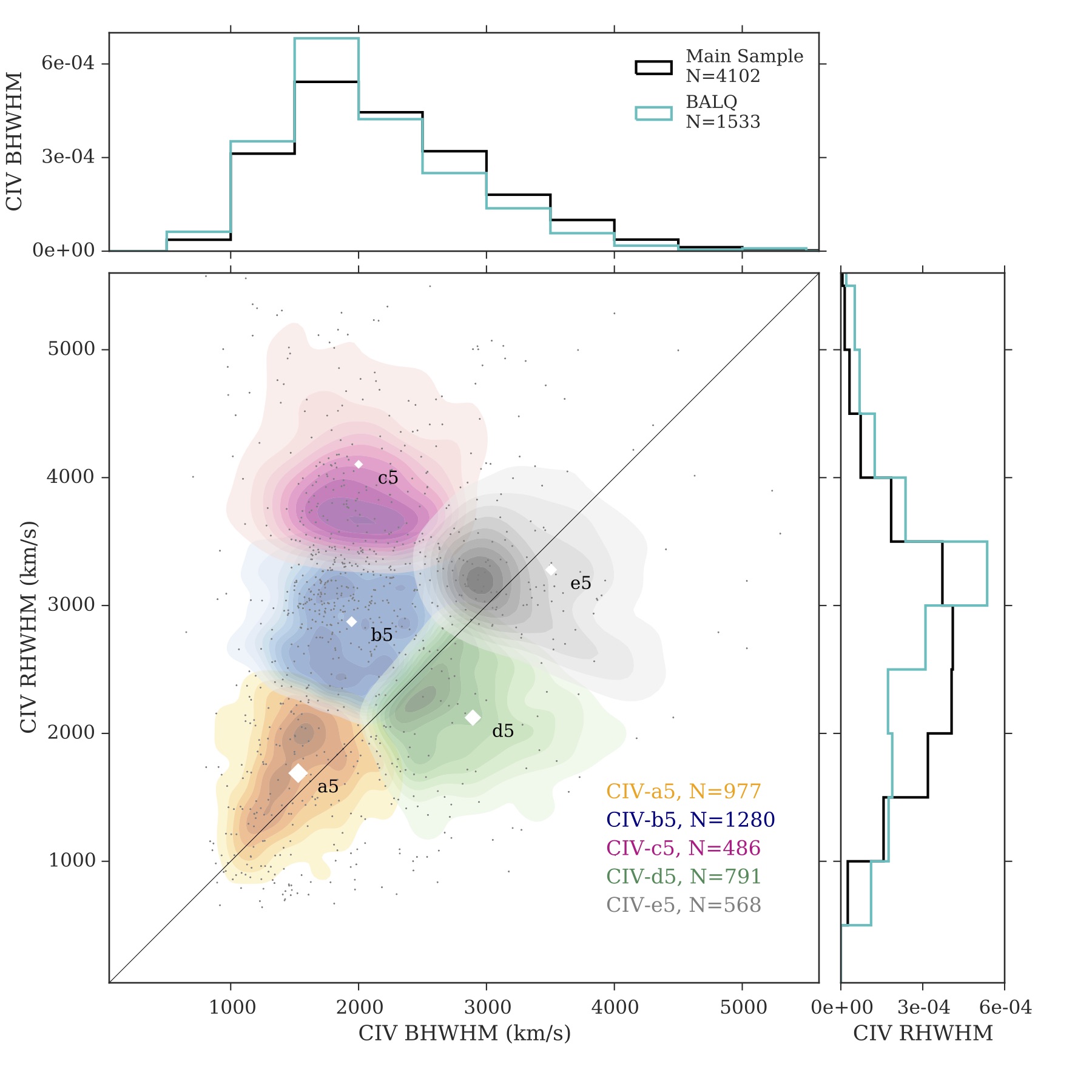}
  \caption{Red- vs. Blue-HWHM of \civ\ for the $K=5$ clustering. We overplot the BAL quasars as grey dots and show the fractional distributions of the \civ\ blue- and red-HWHM of the BALQs and the full sample in the upper and left panels respectively. 
  The distribution of the red-HWHM in the BALQ sample (right panel) peaks at higher values than that of the main sample indicating the red-HWHM of \civ\ is larger in most BAL quasars compared to that of non-BAL quasars.
  We note that the BALQ sample contains many objects with measured HWHM of -1. 
  We discuss these objects further in \S \ref{sec:mixed} and \ref{sec:bal} and show that the algorithm was able to isolate them from the rest of the objects in the sample.}
  \label{fig:c4_k5_kde_bal_hist}
  \end{figure*}

We repeat the same analysis done in (\S \ref{sec:mg2}) but this time using the \civ\ parameters as the features fed to K-means with $K$ =3 and up to 6. 
In Fig. \ref{fig:c4_k5_kde_bal_hist}, we show the clusters for the $K=5$ case with the BAL quasars over-plotted and the distributions of their blue- and red-HWHM compared to the distributions of the main sample. 
The \civ\ clusters are significantly more compact than those of \mgii\ with a dynamical range of $\sim$ 1000--5000 \kms\ for the blue- and red-HWHM and the average \civ\ EWs are larger in objects with narrower \civ\ (Table \ref{tbl:num_clstrs}).

The blue-HWHM of the main and BALQ samples appear to follow similar distributions (top panel in Fig. \ref{fig:c4_k5_kde_bal_hist}) but the distribution of the red-HWHM in the BALQ sample (right panel of Fig. \ref{fig:c4_k5_kde_bal_hist}) is skewed towards higher values than that of the main sample, i.e., the red-HWHM of \civ\ is larger in most BAL quasars than that of non-BAL quasars.
We point out, however, that the \civ\ emission-line measurements for BAL quasars can be sometimes unreliable because the BAL features often interfere with the emission-line fits as the absorption can in many cases be fully or partially superimposed on the emission-line which undermines the measurements of the emission line.

\begin{figure*}
\includegraphics[width=\linewidth]{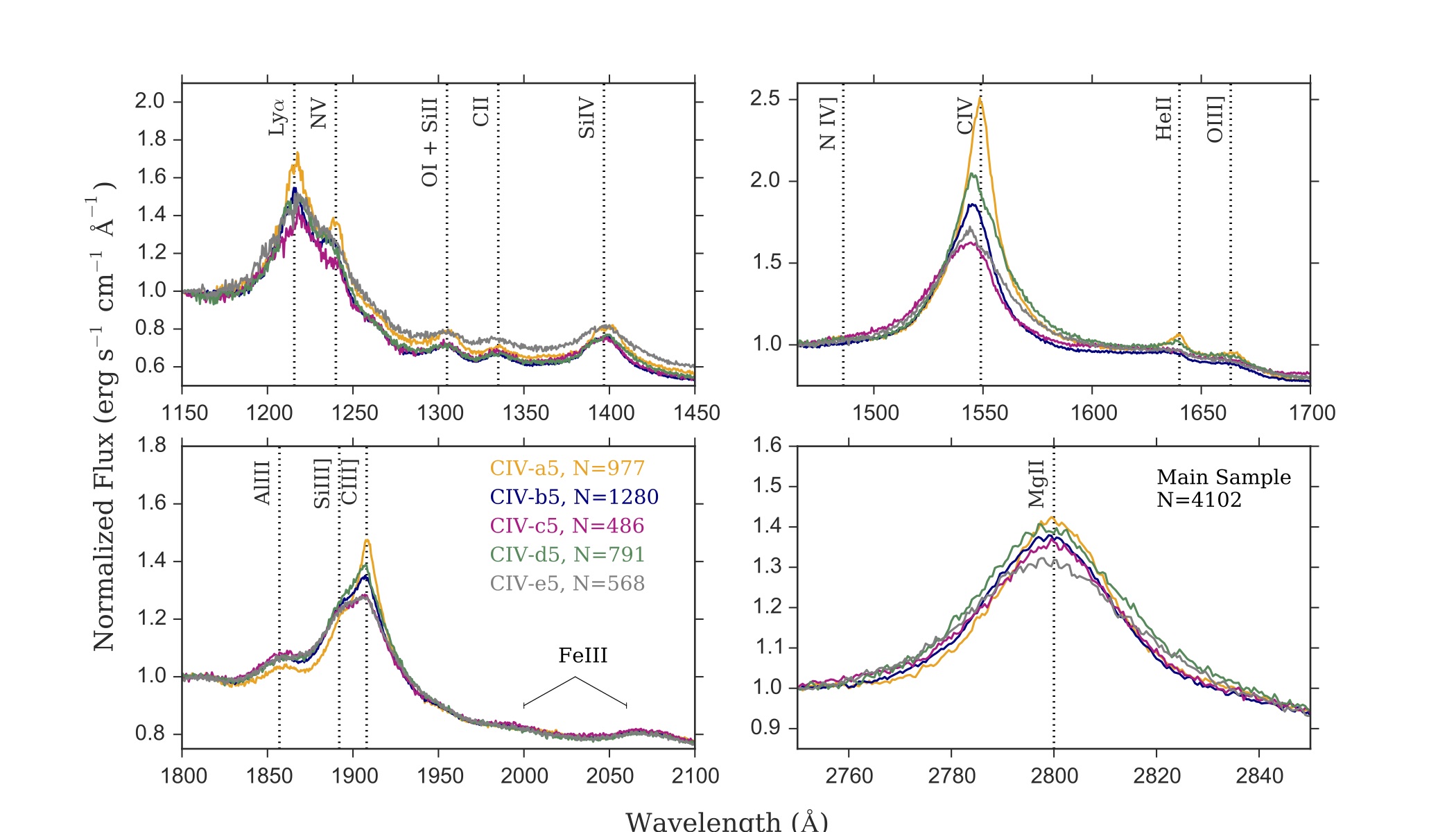}
\caption{Median composite spectra made from the main sample objects in the \civ\ clusters with $K=5$.
 The profiles have been normalized locally at the starting wavelength of each panel. 
 The labels in the lower-left panel refer to the number of objects in each composite/cluster (see also Table \ref{tbl:num_clstrs}).
 \civ\ profiles shift from a large EW, strongly peaked and symmetric line (\civ-a5) to a broader, weaker and blueshifted one (\civ-c5) at the extremes. 
 \civ-a5 also shows narrow, peaked profiles in \lya\ and \ciii\ with a low \siiii/\ciii\ ratio while \civ-c5 has a higher \siiii/\ciii\ ratio.
 The prominence of \heii\ and the low \siiii/\ciii\ ratio in \civ-a5 are signs of a hard ionizing continuum relative to \civ-c5.
 \mgii\ lines have nearly similar widths regardless of the shape of the \civ.
 The weak absorption seen in the \civ\ profiles on \civ-d5 and \civ-e5 is discussed in the text (\S \ref{sec:c4} and Fig. \ref{fig:c4_ex}).}
	\label{fig:c4_k5_profiles}
\end{figure*}

Notably, Fig. \ref{fig:c4_k5_profiles} shows that despite using \civ\ properties \emph{only} to generate the clusters, the other emission lines in the composite spectra generated from the clusters are corresponding to \civ\ in a way that is expected from observations of quasar UV spectra.
For example, composite \civ-a5 in Fig. \ref{fig:c4_k5_profiles} has narrow, symmetric \civ\ which shows no shift from the systemic redshift and has a smaller \siiii/\ciii\ ratio and a prominent \heii\ (also strong \siiv, OI and SiII and narrower \mgii), while the composites with blueshifted and weaker \civ\ (e.g., composite \civ-c5) have larger \siiii/\ciii\ ratios and weaker \heii.
We point out here that the presence of a strong \heii\ line coupled with a higher \siiii/\ciii\ ratio is an indication of a harder ionizing SED -although other interpretations of a high \siiii/\ciii\ ratio alone (such as a higher density) are also possible \citep[][]{leighlymoore04, casebeer06}.
These observations are in line with the disk-wind model of the broad-lines in quasars which predicts that blueshifted high ionization lines are a result of a fast-moving wind emerging from the accretion disk and accelerated outward by radiation pressure, while intermediate and lower ionization lines are emitted closer to the base of the wind at the accretion disk and are therefore seen to be symmetric and at systemic redshift \citep[][]{leighly04, leighlymoore04, richards11}.
The shape of the ionizing SED in the disk-wind model plays a major role in the contributions of each of the disk and wind components to the broad emission lines; quasars with higher UV luminosity relative to the X-ray have a stronger wind component (more blueshift in \civ), while quasars with harder X-ray over-ionizes the gas and have therefore less contribution from the wind to the broad lines \citep[e.g.,][]{casebeer06, leighlymoore04, leighly07, kruczek11, richards11}.

\begin{figure}
\includegraphics[width= \linewidth]{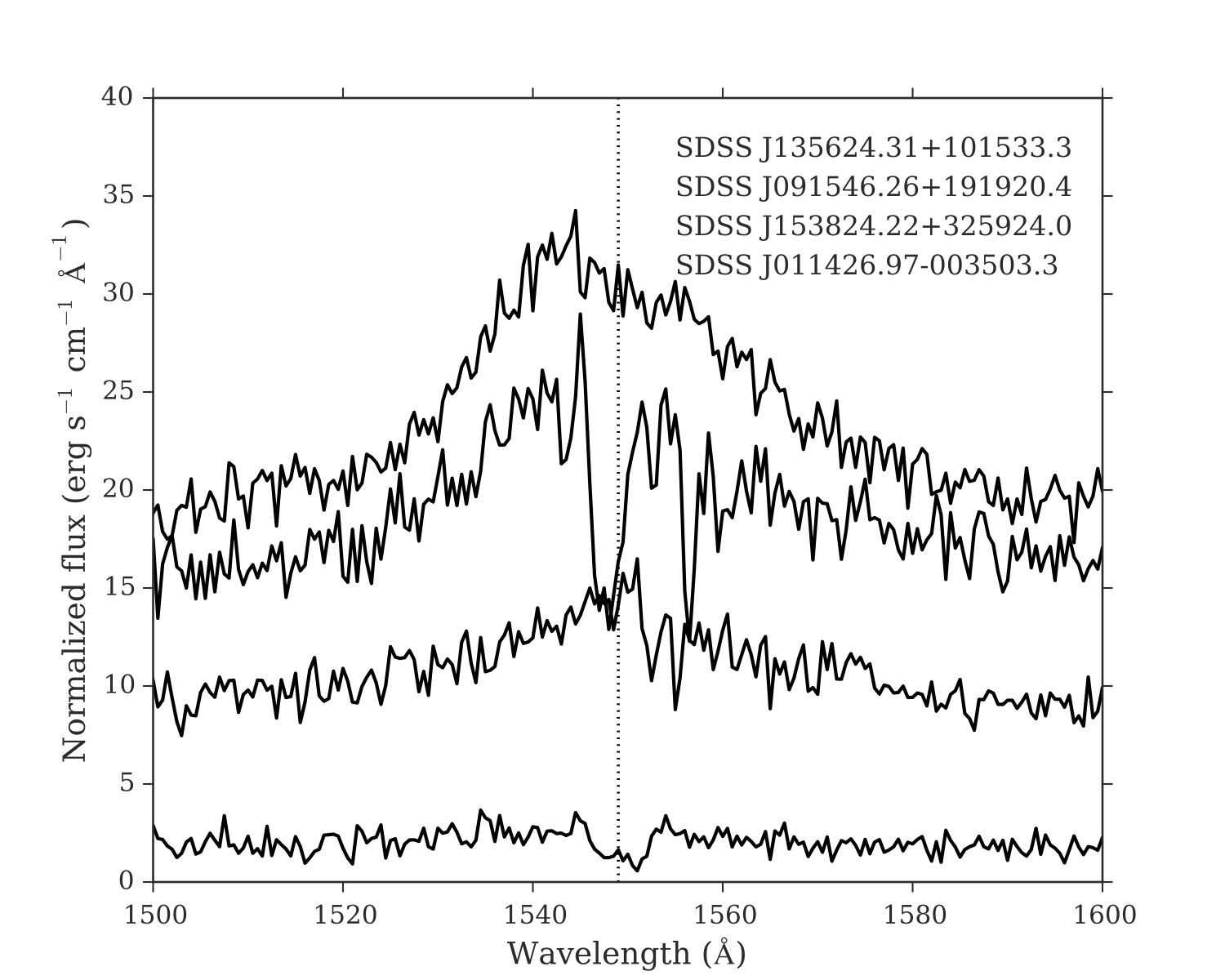}
\caption{Example spectra of individual objects from the \civ-d5 and \civ-e5 composites in the main sample (Fig. \ref{fig:c4_k5_profiles}).
Objects in these clusters have either highly blueshifted \civ\ and/or have absorption that is missed from the BAL visual inspection or is not broad enough to be classified a BAL quasar.}
\label{fig:c4_ex},
\end{figure}

We also see in Fig. \ref{fig:c4_k5_profiles} that clustering done using \civ\ properties results in \mgii\ profiles that are rather similar regardless of the \civ\ profiles (similar to what we see in the \mgii\ clusters in Fig. \ref{fig:mg2_k5_profiles}).
This again cautions that estimates of black hole masses using \civ\ could result in values that do not necessarily reflect the ones estimated by the less biased \mgii.
We finally point out that the weak absorption seen in the \civ\ profiles of composites \civ-d5 and \civ-e5 might indicate a higher number of narrow absorption among objects in these two clusters.
This feature is unlikely to be due to low numbers as cluster \civ-c5 has fewer objects but its \civ\ profile shows no sign of absorption. 
Fig. \ref{fig:c4_k5_kde_bal_hist} shows that these two clusters include objects with extremely high BHWHM. 
We visually examine the spectra of objects from clusters \civ-d5 and \civ-e5 with the highest BHWHM and find that most of these objects have either  absorption in their \civ\ profiles that is not flagged as BAL or have \civ\ profiles that are highly skewed blue-ward (could be a blueshift or absorbed flux on the red side).
Table \ref{tbl:redshifts} indeed shows that the \civ\ redshifts in these two clusters slightly diverge from the PCA redshifts.
In Fig. \ref{fig:c4_ex} we show examples of \civ\ profiles for four objects with extremely high BHWHM in clusters \civ-d5 and \civ-e5.
We see no clear trend in these two clusters with respect to the properties their SED as seen by the \heii\ and the \siiii/\ciii\ ratio in the corresponding composites.

\subsection{\ciii\ Clusters}
\label{sec:c3}

\begin{figure*}
  \includegraphics[width= \linewidth]{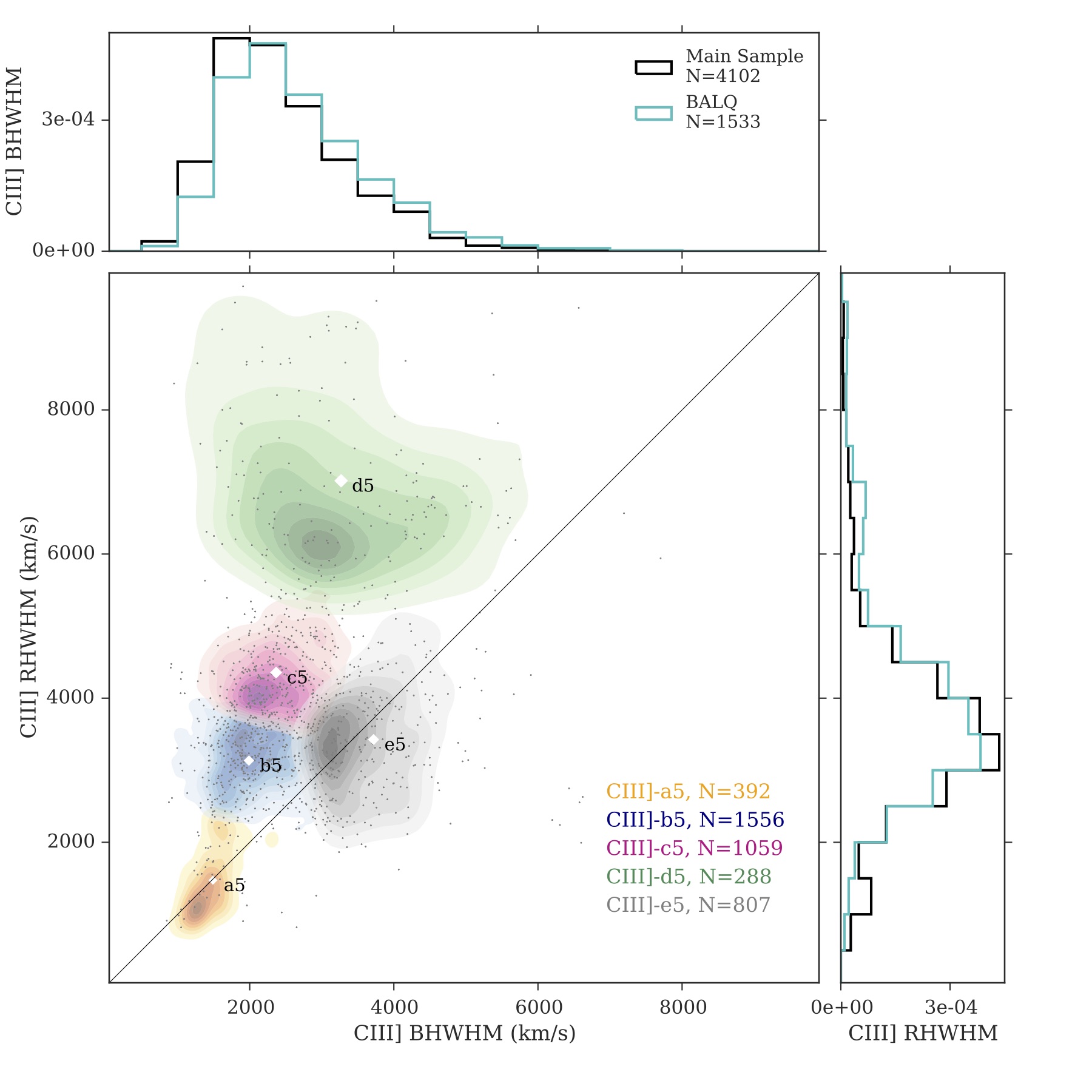}
  \caption{Red- vs. blue-HWHM of \ciii\ for the $K=5$ clustering in the main sample. 
  The diagonal black line marks to 1:1 ratio.
 The white diamonds mark the projected locations of the cluster centroids and their sizes correspond the the average EW of the \ciii\ blend of objects in each cluster which appear to be consistent among clusters.
  We over-plot the BAL quasars (not used in the clustering) with grey dots and show the fractional distributions of the \ciii\ blend blue- and red-HWHM of the BAL quasars in the upper and right panels respectively along with those of the main sample.
We see no significant differences in distributions of both samples.
The larger values of the red-HWHM of the \ciii\ blend (cluster \ciii-d5) reflect a lower \ciii\ line peak relative to the other two lines in the blend as becomes clear when we look at the composites (Fig. \ref{fig:c3_k5_profiles}).}
  \label{fig:c3_k5_kde_bal_hist}
  \end{figure*}

\begin{figure*}
\includegraphics[width=\linewidth]{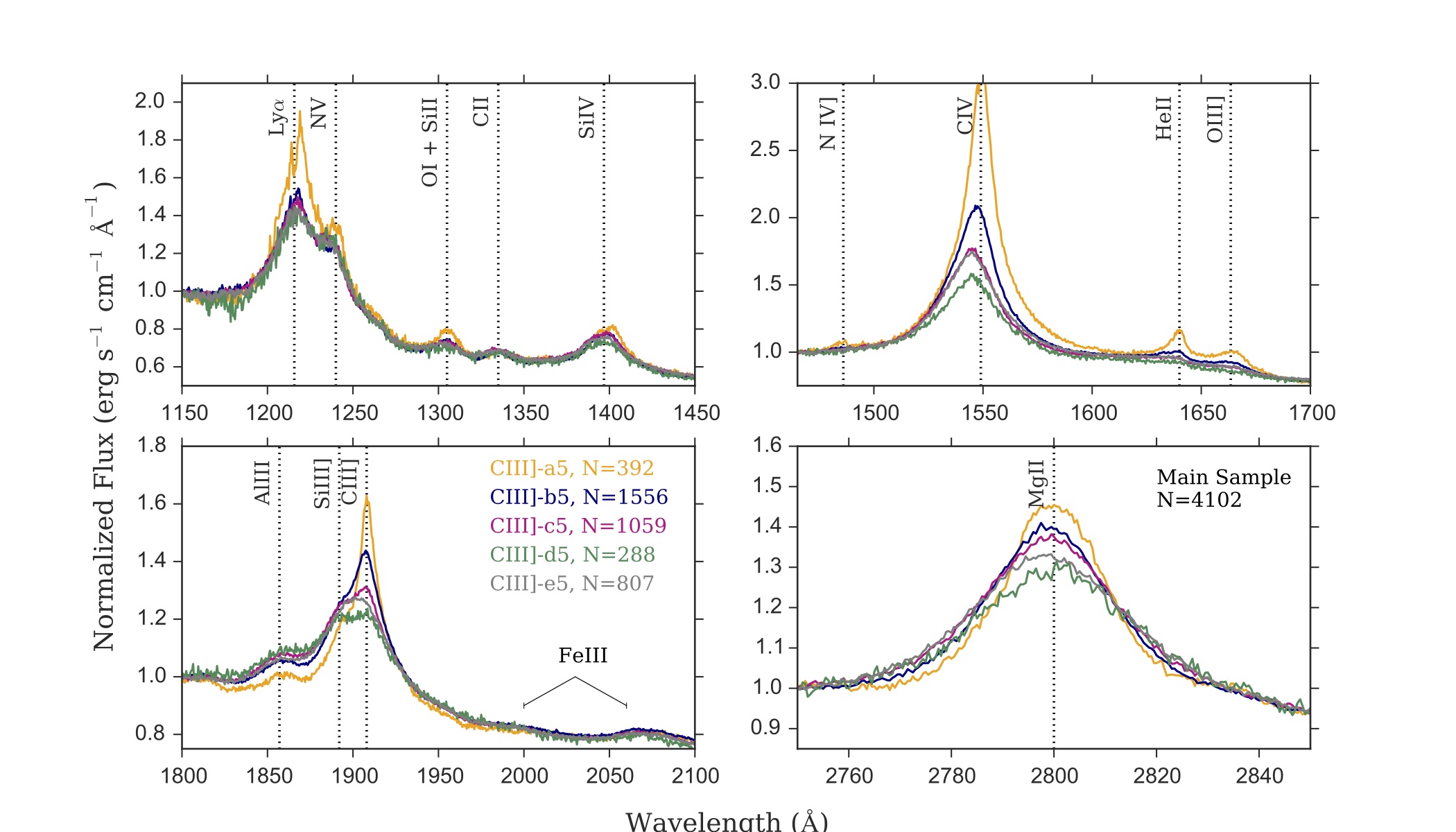}
\caption{Median composite spectra made from the objects in the \ciii\ clusters with $K=5$ shown in Fig. \ref{fig:c3_k5_kde_bal_hist}. 
The profiles have been normalized locally at the starting wavelength of each panel. 
The numbers in the lower-left panel refer to the number of objects in each composite/cluster (see also Table \ref{tbl:num_clstrs}).
Composite \ciii-a5 with large EWs and narrow, symmetric profiles for \civ, \lya, \heii, and \mgii\ has lower \siiii/\ciii\ ratio while composite \ciii-d5 has smaller EW and broader blueshifted \civ\ with higher \siiii/\ciii\ ratio indicating a softer ionizing continuum and a larger wind contribution to the \civ\ profile.}
	\label{fig:c3_k5_profiles}
\end{figure*}

We repeat the clustering analysis here in a similar fashion to the previous two lines and with $K=$ 3, 4, 5, and 6 for the EW, BWHM and RHWHM of \ciii.
Fig. \ref{fig:c3_k5_kde_bal_hist} shows the results of those runs in the BHWHM--RHWHM plane for the $K=$ 5 case.
The figure also shows the distributions of the blue- and red-HWHM of \ciii\ for the main and BALQ samples and the average EW for each cluster.
The blue- and red-HWHM of \ciii\ in BAL and non-BAL quasars are fairly consistent and the EWs of objects in the different clusters do no appear to have a major role in driving the clustering.
Furthermore, the \ciii\ measurements in the \citet{paris14} catalog refer to the \ciii, \siiii, and \aliii\ complex measured without any deblending.
This means that the EW of this blend can be similar in two objects but with significantly different contributions from the lines in the blend.
The red- and blue-HWHM in the \ciii\ blend in this catalog are measured from the line centroid set at the location of the line peak (maximum intensity) of the blend \citep{paris12,paris14}.
This means that objects with extremely high RHWHM in Fig. \ref{fig:c3_k5_kde_bal_hist} are simply objects with weaker \ciii\ relative to the other two lines in the blend (e.g., cluster \ciii-d5 vs. cluster \ciii-b5).
This becomes clear when we examine the composite spectra in Fig. \ref{fig:c3_k5_profiles} generated from median-combining objects in each cluster in Fig. \ref{fig:c3_k5_kde_bal_hist}.
A great advantages of using K-means is the ability to use the \ciii\ blend as a whole (the entire line blend) to isolate objects with varying strengths of \ciii, \siiii, and \aliii\ without having to deblend them (which might indeed be challenging with survey quality data).

Again in this case, the clusters made using the \ciii\ complex properties are reflecting similar trends to those found in the composites in Fig. \ref{fig:c4_k5_profiles} and are perhaps even more pronounced.
Indeed objects with large EWs and narrow, symmetric profiles for \civ\ (also \lya, \heii, and \mgii; composite \ciii-a5 in Fig. \ref{fig:c3_k5_profiles}) have a lower \siiii/\ciii\ ratio while objects with lower EWs and broader blueshifted lines (composites \ciii-d5 and \ciii-e5 in the same figure) have a higher \siiii/\ciii\ ratio.
This suite of observations is in agreement with the disk-wind model discuss in \S \ref{sec:c4} in which the shape of the ionizing SED has strong effects on the lines in a sense that a quasar with a soft ionizing SEDs (weak \heii, high \siiii/\ciii\ ratio, for example \ciii-d in Fig. \ref{fig:c3_k5_profiles}) exhibits stronger wind component (weaker \civ\ with larger blueshifts), while a harder SED over-ionizes the wind resulting in a stronger \civ\ that is centred at zero velocity \citep[composite \ciii-a; e.g.,][]{casebeer06, leighly07, richards11}.

Clusters \ciii-a5 through \ciii-d5 in Fig. \ref{fig:c3_k5_kde_bal_hist} appear to follow a sequence with increasing red- and blue-HWHM --a trend that is also reflected in their \ciii\ profiles in Fig. \ref{fig:c3_k5_profiles}.
Cluster \ciii-e5 seems however to fall off this sequence with its red-HWHM similar to those of \ciii-b5 and \ciii-c5 but with larger blue-HWHM.
Its composite spectra in Fig. \ref{fig:c3_k5_profiles} shows that the line profile also falls off the sequence of the decreasing \ciii/\siiii\ ratio from \ciii-a5 through \ciii-d5.
In addition, the \civ\ profile in this cluster does not follow the gradual increase in blueshift from \civ-a5 to \civ-d5 and its \mgii\ appears to be slightly skewed blue-ward systemic.
Table \ref{tbl:redshifts} shows that the \ciii\ redshift estimates for this cluster are significantly different from those of PCA ($p$ value $\simeq 0.1$) which potentially indicates discrepancy in the blue- and red-HWHM of the \ciii\ blend in this cluster due to problematic redshift estimates.
Table \ref{tbl:redshifts}  also shows that \civ\ redshifts for this cluster are in less agreement with the PCA redshifts.

Another notable feature in Fig. \ref{fig:c3_k5_profiles} is the relatively strong N IV] 1486 \AA\ line which becomes prominent in composite \ciii-a5.
The low \siiii/\ciii\ ratio and the strong \heii\ in this composite indicate a harder ionizing SED.
This nitrogen feature is rather uncommon in quasar spectra --\citet{bentz04} found less than 1\% of their SDSS sample has enhanced nitrogen and our visual examination of selected objects from cluster \ciii-a5 confirms the scarcity of N IV] 1486 \AA\ in this cluster.
\citet{jiang08} studied a sample of 293 quasars with strong nitrogen emission lines (N IV] $\lambda1486$ or N III] $\lambda1750$) and found that quasars with higher nitrogen abundance share similar properties with the overall quasar population though higher nitrogen abundance is associated with higher fraction of radio-loud objects.
We look at the fraction of objects detected by FIRST and find that indeed, cluster \ciii-a5 includes a slightly higher fraction of quasar detected by FIRST (12.8 \%) compared to the other clusters which contain a range of 7.1\% to 8.7\% FIRST-detected objects.

\subsection{Clustering on the Mixed Sample}
\label{sec:mixed}

Next, we set the BAL flag to 1 in the selection process mentioned in \S \ref{sec:selection}, allowing BAL quasars to be included in the sample.
We also remove the constraint that EW $>0$ and keep the redshift limit (1.6--2.1) and the S/N and EW uncertainty cutoffs: S/N($\lambda1700) >3$, EW error $<10 \%$ of measured EW.
Adding BAL quasars to the sample increased its size to 6463 (not including the 15 objects we previously removed from the main sample).
The purpose of defining this mixed sample is to examine the \ciii\ properties of the entire quasar population (BAL and non-BAL quasars).
We do not use \civ\ and \mgii\ measurements for the clustering analysis here because (by definition) at least one of these lines shows absorption in BAL quasars and can therefore bias the results.

\begin{figure*}
\includegraphics[width=\linewidth]{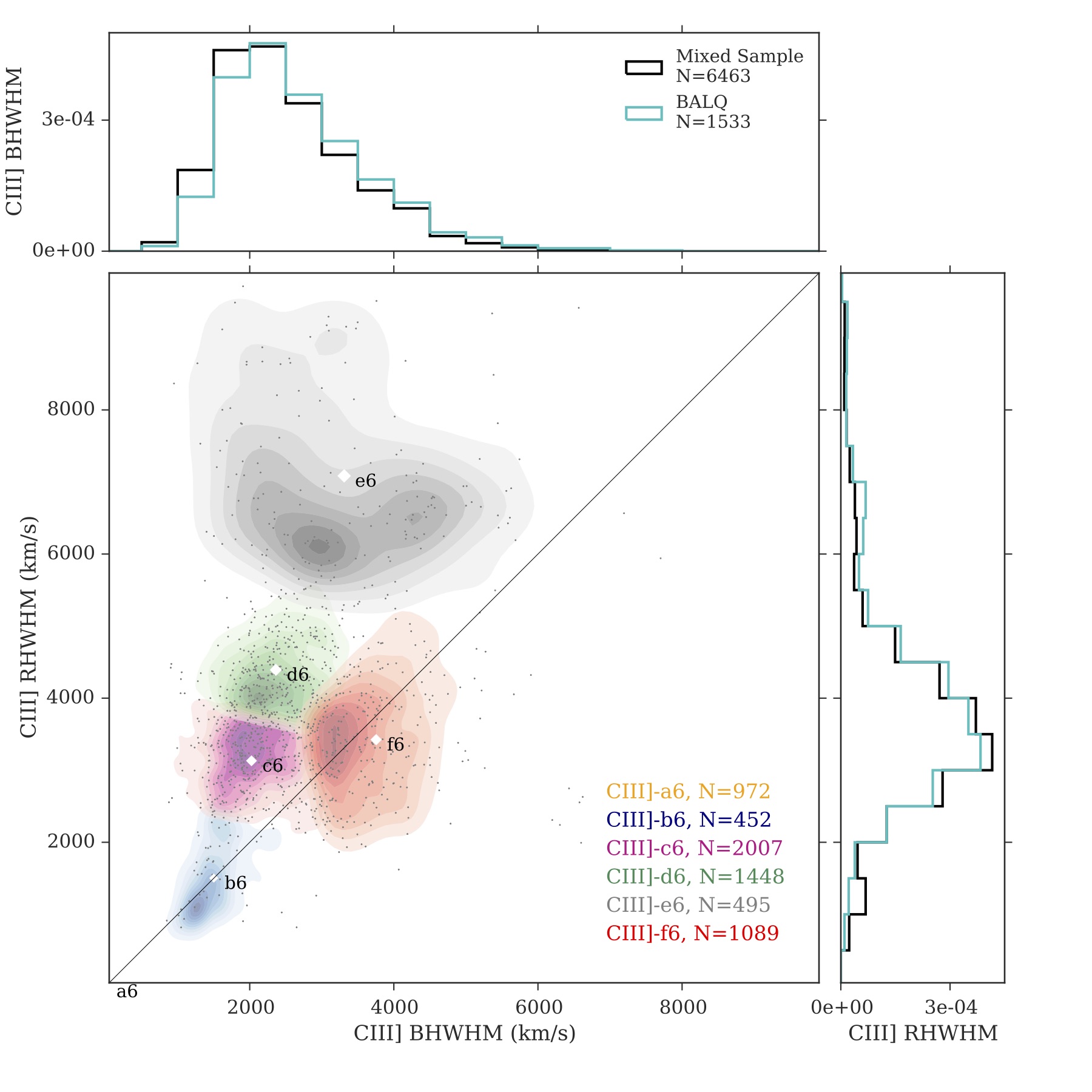}
\caption{The red- vs. the blue-HWHM of the \ciii\ blend in the mixed sample for the clustering done using $K=$ 6.
The white diamonds show the projected position of the cluster centroids and their sizes correspond to the the average EW of the \ciii\ blend in each cluster.
Cluster \ciii-a6 (containing objects with failed fitting) is not shown.
The BAL quasars (included in this sample) are over-plotted as grey dots and the histograms show the fractional distributions of the BHWHM and RHWHM in the mixed and the BALQ samples and show no significant difference.
The fraction of BAL quasars varies among the clusters: \ciii-a6 (33 \%), \ciii-b6 (11 \%), \ciii-c6 (19 \%), \ciii-d6 (24 \%), \ciii-e6 (33 \%), \ciii-f6 (23 \%).
Those differences in the BLAQ fractions potentially reflect differences in the hardness of the SED (see \S \ref{sec:mixed} and Fig. \ref{fig:c3_k6_mixed_profiles}).
}
\label{fig:c3_k6_kde_mixed_hist}
\end{figure*}

\begin{figure*}
\includegraphics[width=\linewidth]{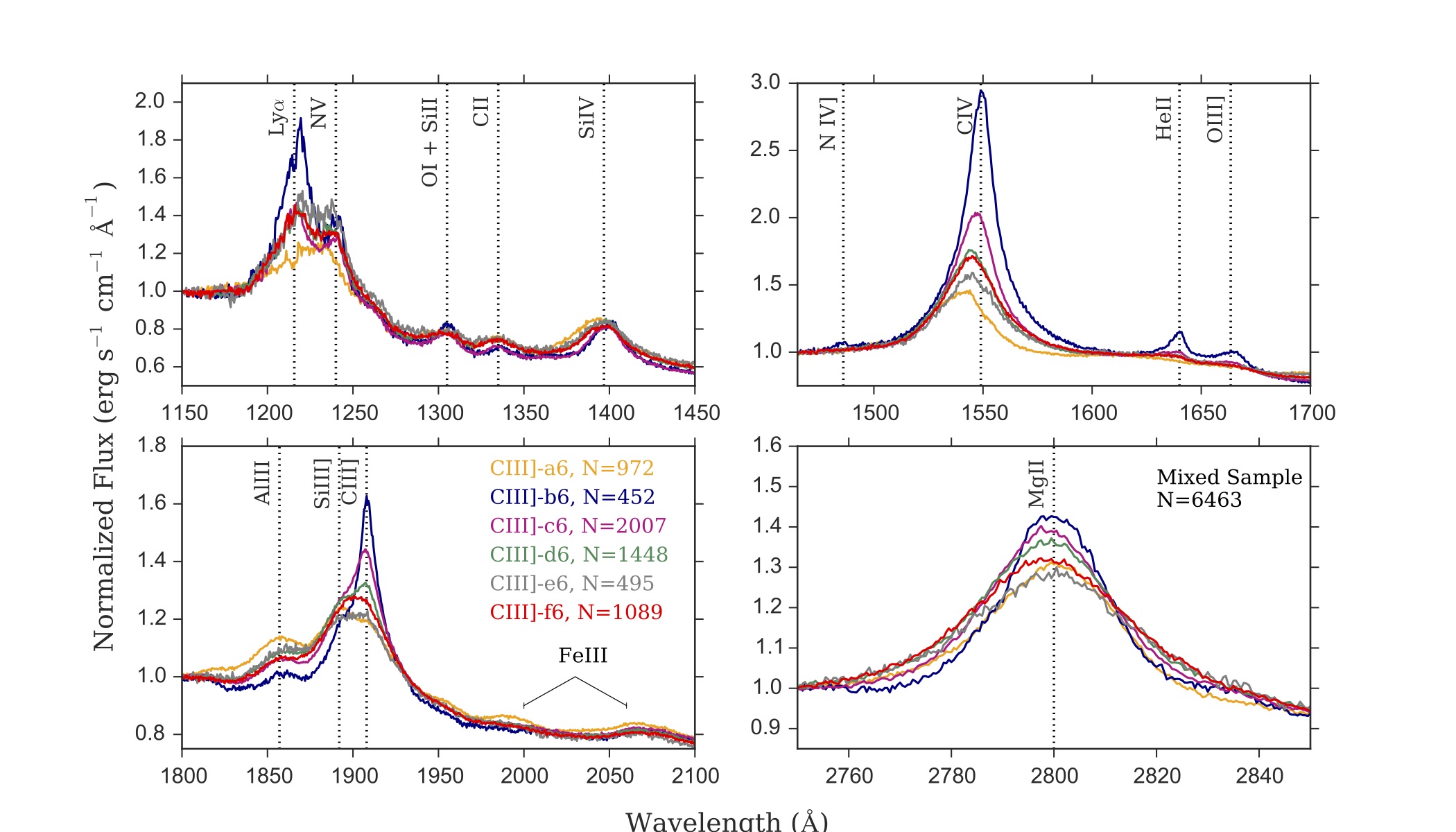}
\caption{Median composite spectra made from the objects in the \ciii\ clusters shown in Fig. \ref{fig:c3_k6_kde_mixed_hist} with $K=6$ in the mixed sample. 
The profiles have been normalized locally at the starting wavelength of each panel and the numbers in the lower-left panel refer to the number of objects in each composite (cluster).
Composite \ciii-b6 has prominent narrow features such as \civ, \lya, and \ciii\ and weak \aliii\ and \siiii\ (low \siiii/\ciii\ ratio), while composites \ciii-e6 and \ciii-f6 have weaker \civ\ that is blueshifted and stronger \siiii/\ciii\ ratios.
Composite \ciii-a6 is created from the objects with the failed fitting (-1 for the \ciii\ EW, RHWHM and BHWHM as reported in the catalog).
The relatively strong Fe III feature in this cluster might indicate the weak-line quasars comprise a large portion of objects in this clusters.}
\label{fig:c3_k6_mixed_profiles}
\end{figure*}

Fig. \ref{fig:c3_k6_kde_mixed_hist} shows the clusters using the \ciii\ blend properties in the mixed sample for $K=$ 6 run.
We find that in the mixed sample, the algorithm is separating a group of objects that were not fit in the \citet{paris14} catalog and a "-1" was entered instead of the measurements of the red- and blue-HWHM and EW.
It is only when $K=6$ that cluster \ciii-a6 is almost purely made of the ``negative'' objects (971 objects and only one object with non-negative values (22.6 \AA, 741 \kms, 695.5 \kms)).
Of these objects in cluster \ciii-a6 $\sim 33\%$ are flagged as BAL quasars in the \citet{paris14} catalog while the rest are not marked as BAL quasars.
Visual inspection of the spectra of individual objects in cluster \ciii-a6 shows that many of them have ``flat'' spectra (no emission lines and in some cases what looks like weakly absorbed lines) and so the automated fitting routine was not able to find lines to fit either because the lines are very weak or because of the strong BAL troughs.
Objects with BAL troughs could be easier to misfit as the absorption troughs could be highly blueshifted and cause the fitting routine to fail.
Some of the objects in cluster \ciii-a6 with the weak emission lines might potentially meet the definition of the so-called weak-lined quasars \citep[e.g.,][and references therein]{luo15,shemmer15}.
This is supported by the presence of a more prominent Fe III feature blue-ward of \ciii\ known to be stronger in weak-lined quasars as can be seen in Fig. \ref{fig:c3_k6_mixed_profiles} which shows  the composite spectra generated from objects in each of the clusters.

It is also intriguing to see that the objects collected in cluster \ciii-a6 (with the -1 values) are generating a weak, strongly blueshifted \civ\ profile which may contain absorption features as can be inferred from the dip in \lya.
This weak \civ\ line is also associated with the highest \siiii\ to \ciii\ ratio indicating a soft ionizing SED on average in those objects compared to the other clusters.
Similar to what we see in the main sample (\S \ref{sec:c4} and \S \ref{sec:c3}), the cluster with the narrow, strong \civ\ line which is centered at the systemic redshift (composite \ciii-b6) has the strongest \heii\ and the smallest \siiii\ to \ciii\ ratio consistent with a hard ionizing continuum which inhibits wind formation.

Finally, it is worth mentioning that the fraction of BAL quasars in each cluster appears to support the notion that BAL quasars are more likely to have softer (X-ray weaker) SEDs as we find that the numbers of BAL quasars in each of the mixed cluster decreases gradually with the \siiii/\ciii\ ratio; cluster \ciii-b6 in Fig. \ref{fig:c3_k6_mixed_profiles} (with the strongest \heii\ and lowest \siiii/\ciii\ ratio) has only 11\% BAL quasars while clusters \ciii-a6 and \ciii-e6 (with relatively weaker \heii\ and higher \siiii/\ciii\ ratio) have each 33\% of its objects flagged as BAL quasars in the catalog.
The percentage in the rest of the clusters is: \ciii-c6 has 19\% BAL quasars, \ciii-d6 has 24\% BAL quasars and \ciii-f6 has 23\% BAL quasars.

\subsection{Sample with BAL Quasars Only}
\label{sec:bal}

We now look at the properties of 1533 BAL quasars selected as described in \S \ref{sec:mixed} but with the BAL flag set to 1.
Here, too, the algorithm is separating the objects with no fitting in the catalog (with -1 entreis for the EW and widths).
Up to $K=6$, the cluster includes 8 objects with $> -1$ values with the remaining 327 objects with -1 for the measurements of \ciii\ EW and blue- and red-HWHM.

\begin{figure*}
\includegraphics[width=\linewidth]{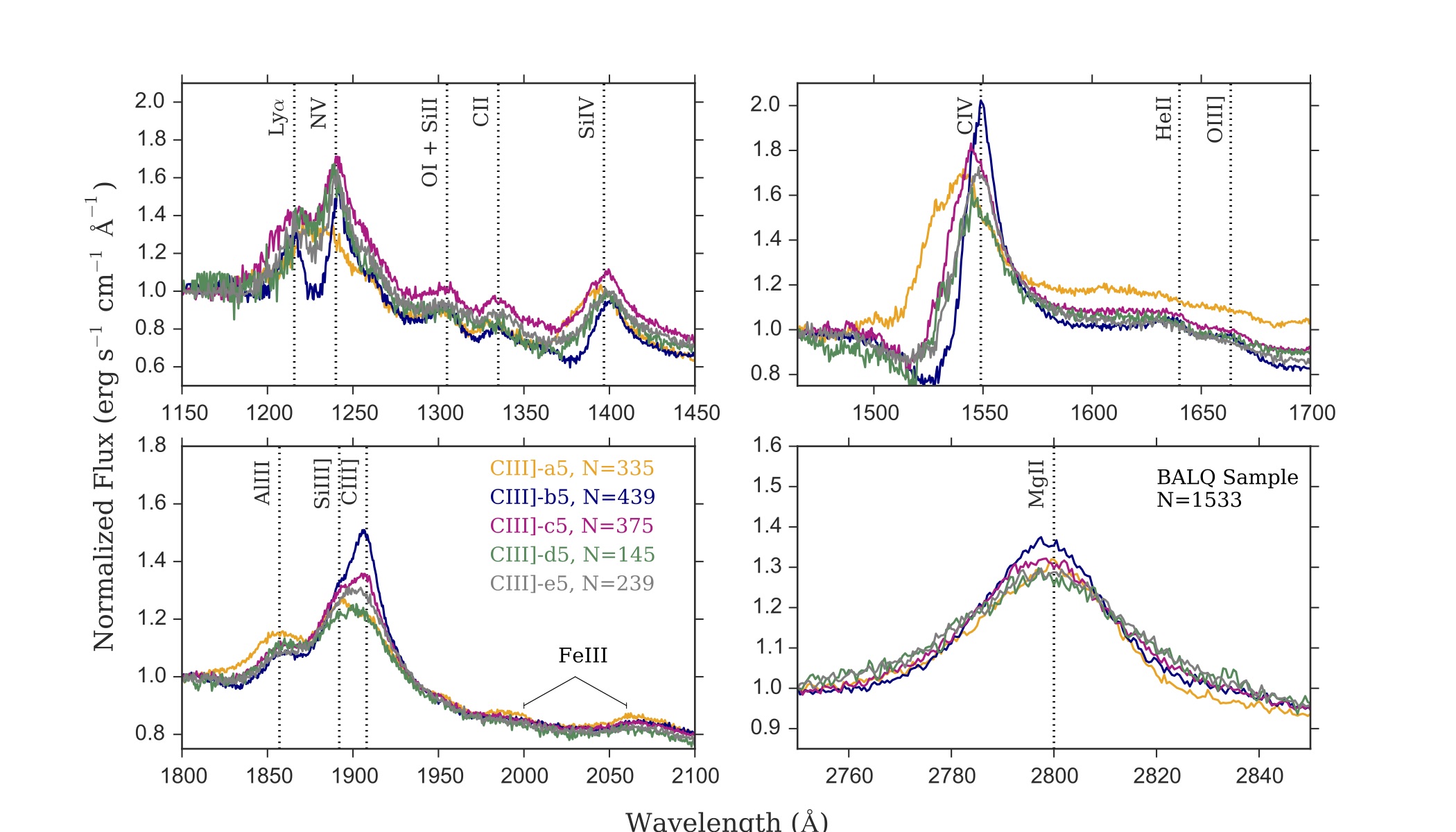}
\caption{Median composite spectra made from the objects from the BALQ sample in the \ciii\ clusters with $K=5$. 
The profiles have been normalized locally at the starting wavelength of each panel and the numbers in the lower-left panel refer to the number of objects in each composite (cluster).
Composites made from clusters of the \ciii\ blend properties appear to have different emission and absorption in \civ\ and \siiii, e.g., 
composite \ciii-b5 has the lowest \siiii/\ciii\ ratio and a narrow \civ\ emission-line that is not blueshifted and a low V$_{min}$ BAL though, 
while composites \ciii-d5 and \ciii-e5 have a higher ratio and are both blueshifted in their \civ\ emission and their BAL troughs have higher V$_{min}$.
Composite \ciii-a5 mostly contains objects with -1 measurements for the \ciii\ blend and it shows a weaker \civ\ emission and a shallow BAL trough with the highest V$_{min}$.
\siiv\ shows similar trends to those seen in \civ.}
	\label{fig:c3_k5_bal_profiles}
\end{figure*}

Figure \ref{fig:c3_k5_bal_profiles} shows the median composite spectra generated from objects in the $K=$ 5 run.
We find that the \civ\ profiles (both in emission and absorption) are changing with the properties of the \ciii\ blend used to generate the clusters.
Composite \ciii-b5, for example, with the lowest \siiii/\ciii\ ratio, has a strong, narrow \civ\ emission-line centred at the systemic redshift and a deep absorption trough with low V$_{min}$. 
On the other hand, composites with larger \siiii/\ciii\ ratios (\ciii-d5 and e5) have blueshifted \civ\ emission-lines and higher V$_{min}$ for their absorption troughs.
Composite \ciii-5a --mostly comprised of objects with the -1 measurements of \ciii-- has the highest \siiii/\ciii\ ratio and shows weak, blueshifted \civ\ emission-line and a shallower trough and a very high V$_{min}$.
The absorption troughs in \siiv\ appear to follow similar trends of those seen in \civ.
The lack of any absorption showing in the \mgii\ profiles in those composites indicates that objects in these clusters contain either only high ionization BAL quasars or a mixture of both high and low ionization BAL quasars that averages out in the \mgii\ profiles.

In future work (Tammour et al., in prep), we explore the properties of the \ciii\ complex and the absorption lines in detail for a sample of BAL quasars. 

\section{Caveats}
\label{sec:caveats}
The K-means algorithm is an unsupervised clustering method.
This means that, by definition, there are no fixed labels to the objects in the clusters defined by K-means and it is the investigator's judgement whether or not to accept the clusters found by the algorithm. 
The two main decisions to be made beforehand are what features to use to define the parameter space, and how many clusters to group the objects into.
Deciding which features to use is limited to what is available and is strongly related to the problem in hand.
In this work, we use the EW and the blue- and red-HWHM of individual lines or line blends.
We decide to use these three parameters as they adequately represent the relative strength and structure (asymmetry) of the emission lines.
Determining $K$ (the number of clusters), on the other hand, is a heuristic exercise.
It requires familiarity with the dataset and the scientific question in hand, and a few iterations of choosing a $K$ value and examining the output.
In our case, we find that even though $K=3$ gives nicely separated clusters, increasing $K$ to 5 or 6 allows us to isolate some of the outliers that are otherwise ``hiding'' in the clusters.
The two metrics we use (discussed in \S \ref{sec:kmeans}) are heuristic and are useful for guidance --this is why we opt to test a range of K values rather than one single value.
In choosing $K=5$ (and $K=6$ for the mixed sample) we aim to balance using a low number of clusters that could potentially hide interesting properties of the objects and fracturing the cluster into smaller bits that might not show interesting properties.

Finally, we note that results based on machine learning techniques strongly rely on the input data fed to the algorithm.
For this work we use measurements from the Quasar properties catalog of \citet{paris14}.
This catalog includes results of automated measurements of 166,583 quasars and cases where the fitting routine failed to find a proper fit or where broad absorption was not properly flagged are not surprising.
One particularly useful outcome of using K-means is its ability to isolate objects with special features from a large sample without prior knowledge of their peculiar properties.
An example of this is isolating a group of objects (cluster \ciii-e5, Figures \ref{fig:c3_k5_kde_bal_hist} and \ref{fig:c3_k5_profiles} in \S \ref{sec:c3}) with redshifts that are significantly different from those determined using other lines and the PCA redshift determined with the overall fit (see also Table \ref{tbl:redshifts}).
Indeed one of the challenges of working with large datasets is identifying outliers that could potentially be interesting cases or simply inaccurate measurements that went unnoticed and this rather simple technique offers a straightforward way of identifying such groups.

\section{Summary and Conclusion}
\label{sec:conclusion}
In this work we explore the use of unsupervised clustering analysis in searching for patterns among a large number of quasar UV spectra in a multidimensional space.
The K-means algorithm allows us to find clusters in the EW, BHWHM, and RHWHM parameter space of three different UV emission lines and line blends: \civ~1450\AA, \ciii~1903\AA, and \mgii~2800\AA\ (see \S \ref{sec:kmeans}).
We combine objects in each of the clusters we find in median composite spectra to examine the properties of the emission lines in each cluster.
 As we mention in \S \ref{sec:kmeans}, the parameter space we use in this work is expected to be populated continuously and thus K-means is providing us with a way of grouping objects within the continuum that help to reveal the trends from one extreme to the other.
Indeed, our composite spectra show that the line properties move gradually between two extremes.  
The compelling part of the analysis comes from using one line's strength and asymmetries to probe physical properties of the objects (such as the case for the \ciii\ blend probing the hardness of the SED) and seeing the effects of this on other lines not used in the clustering (such as \civ\ and \heii).

We summarize our findings as follows:
\begin{itemize}
\item K-means is a simple yet powerful algorithm that, instead of binning data at fixed and rather arbitrary boundaries, allows us to more freely explore the structure in a multidimensional parameter space and to find clusters of objects with similar properties in this space.
\item When stacked in composite spectra, the clusters we find show some of the well-known trends in quasar UV spectra such as the correlations among the \civ\ blueshift and its EW and the shape of the ionizing SED probed by the strength of \heii\ and the \siiii/\ciii\ ratio (Figures \ref{fig:c4_k5_profiles}).
\item More interestingly, we find this same inverse correlation in the \civ\ line using emission lines that are seemingly not part of this correlation such as the \ciii\ blend (Fig. \ref{fig:c3_k5_profiles}). 
Because \ciii\ is generally not contaminated by absorption, it is perhaps more useful than \civ\ and \mgii\ for finding like objects in a quasar sample.
\item We find that, unlike \civ\ and \ciii, the properties of \mgii\ are not strongly correlated with those of the other lines in the spectra; the width of \civ\, for example, does not show any clear correlation with that of \mgii.
This could lead to potential discrepancy in the determination of black hole masses using broad lines in single epoch spectra (Fig. \ref{fig:mg2_k5_profiles}).
\item We use this technique to examine the properties of \ciii\ in a mixed sample of BAL and non-BAL quasars and find that the properties of \ciii\ in the mixed sample recovered similar trends to those we find in the main non-BAL quasar sample (Figures \ref{fig:c3_k6_kde_mixed_hist} and \ref{fig:c3_k6_mixed_profiles}).
\item In the mixed sample, we find a higher fraction of BAL quasars in composites with weaker \heii\ and higher \siiii/\ciii\ ratio (softer SEDs), while composites with stronger \heii\ and lower \siiii/\ciii\ ratio (harder SEDs) have a lower fraction of BAL quasars (\S \ref{sec:mixed}, Fig. \ref{fig:c3_k6_kde_mixed_hist}). 
\item We find that a somewhat more prominent NIV]~1486 \AA\ feature is seen in some of the composites generated from the \ciii\ clusters (composite \ciii-a5 in Fig. \ref{fig:c3_k5_profiles}, and composite \ciii-a6 in Fig. \ref{fig:c3_k6_mixed_profiles}). 
The stronger HeII and higher \siiii/\ciii\ ratio in those composites indicate a hard(er) SED and so the presence of stronger NIV] is not surprising  (Karen Leighly, private communication) given that the IP for the NIV]~$\lambda1486$ line is 47.4 eV.
Higher nitrogen abundances have been detected before in radio-loud objects \citep[e.g.,][]{jiang08} and we find a slightly higher fraction of FIRST-detected objects (12\%) in those clusters compare to the ones with weaker nitrogen (7-8\%).
\item When we apply the clustering to a sample of BAL quasars only using the measurements of \ciii, we find evidence that the properties of \ciii\ are able to separate objects into groups with different properties of their \civ\ emission and BAL troughs (Fig. \ref{fig:c3_k5_bal_profiles}).
We investigate this result further in Tammour et al. (in prep).
\end{itemize}

\section*{Acknowledgments}
We thank the anonymous referee for helpful comments that improved the manuscript. 
We are grateful to Patrick Hall and Eric Feigelson for useful discussions at an early stage of this work, Nur Filiz Ak and Nathalie Thibert for comments, and Karen Leighly for thoughtful suggestions that helped improve our manuscript.
We also thank Pauline Barmby for inspiring Figures \ref{fig:mg2_k5_kde_bal_hist}, \ref{fig:c4_k5_kde_bal_hist}, \ref{fig:c3_k5_kde_bal_hist}, and \ref{fig:c3_k6_kde_mixed_hist}.
This work was supported by the Natural Science and Engineering Research Council of Canada, and the Ontario Early Researcher Award Program (A.T., S.C.G.).  
This research made use of Astropy, a community-developed core Python package for Astronomy \citep{astropy13} and Matplotlib: A 2D Graphics Environment \citep{matplotlib}.
Funding for SDSS-III has been provided by the Alfred P. Sloan Foundation, the Participating Institutions, the National Science Foundation, and the U.S. Department of Energy Office of Science. The SDSS-III web site is \url{http://www.sdss3.org/}.

\clearpage

\appendix

\section{Spectra}
\label{sec:spec}
For completeness, we show here the composite spectra for the rest of the $K$ runs that we did not include in the analysis in \S\ref{sec:results} for the main sample as well as the mixed and BLAQ samples.
The median composite spectra are generated in a similar manner to what is described in \S \ref{sec:compos} however the trends discussed in the text are robust to the specific value of $K$ chosen for the clustering.

\subsection{Main sample -\mgii}

\begin{figure*}
\includegraphics[width=\linewidth]{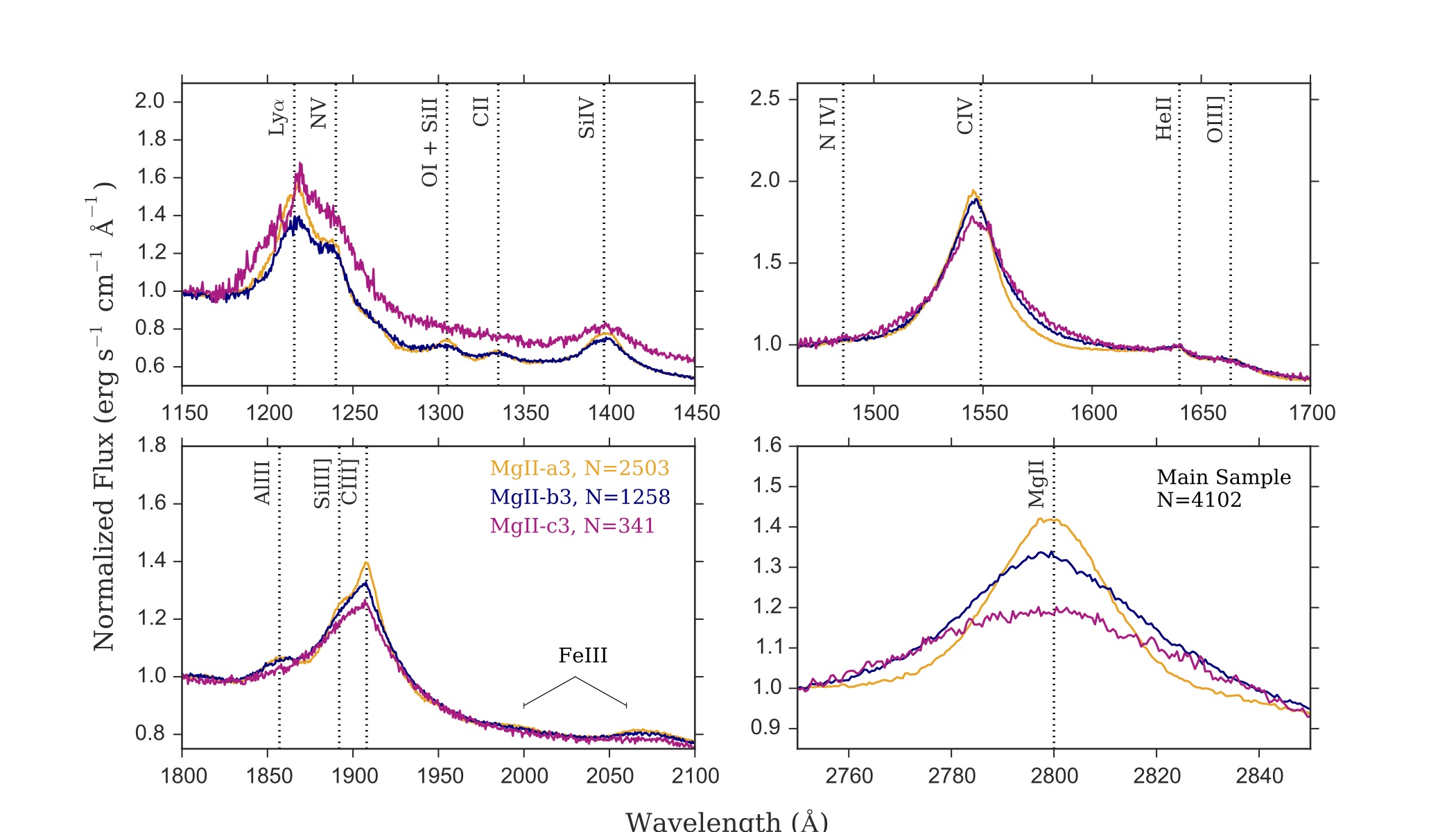}
\caption{Median composite spectra made from the objects in the \mgii\ clusters in Fig. \ref{fig:mg2_clstrs}. 
Similar to Fig. \ref{fig:mg2_k5_profiles} but for $K=3$.}
	\label{fig:mg2_k3_profiles}
\end{figure*}

\begin{figure*}
\includegraphics[width=\linewidth]{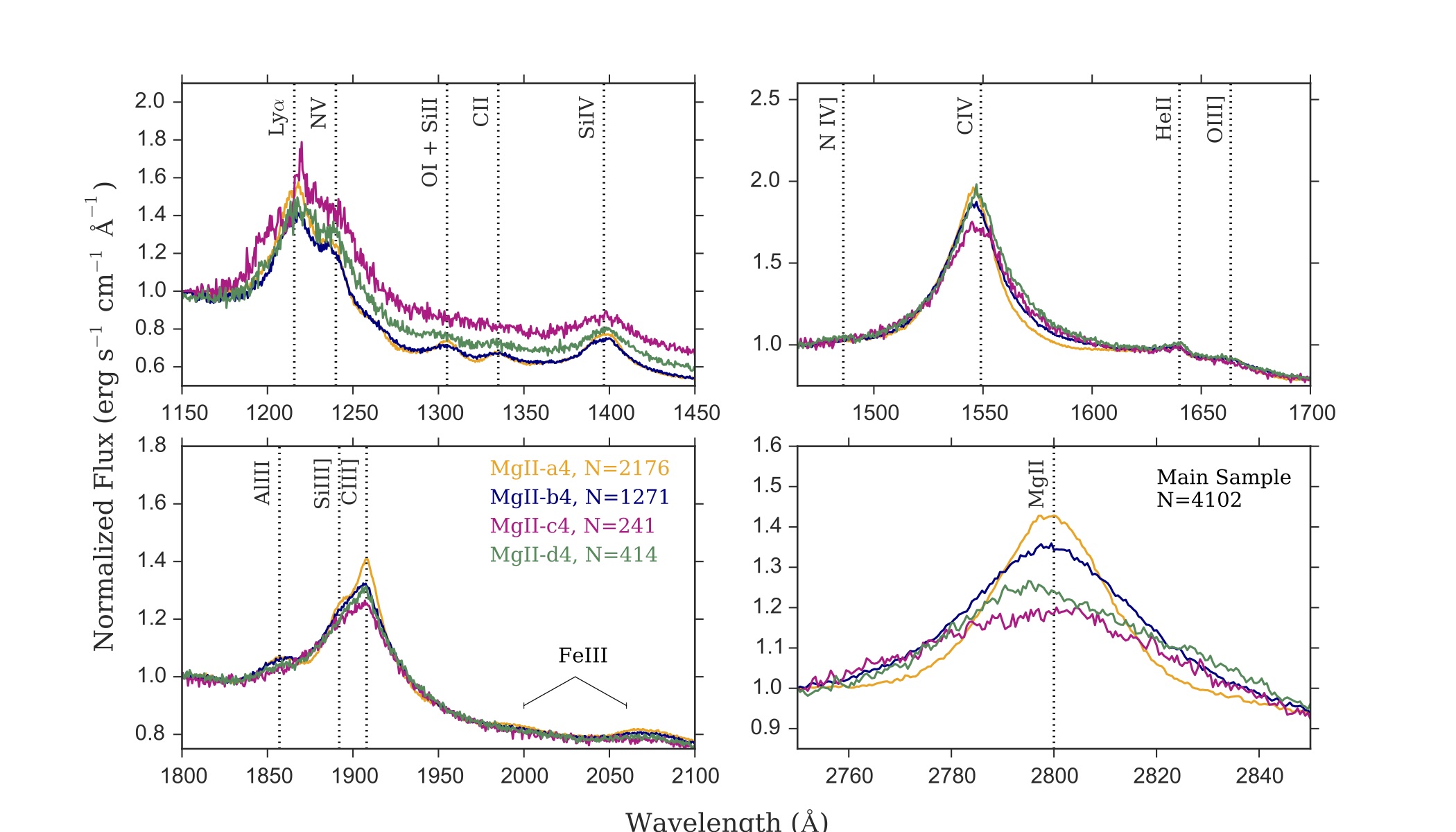}
\caption{Median composite spectra made from the objects in the \mgii\ clusters in Fig. \ref{fig:mg2_clstrs}
Similar to Fig. \ref{fig:mg2_k5_profiles} but for $K=4$.}
	\label{fig:mg2_k4_profiles}
\end{figure*}

\begin{figure*}
\includegraphics[width=\linewidth]{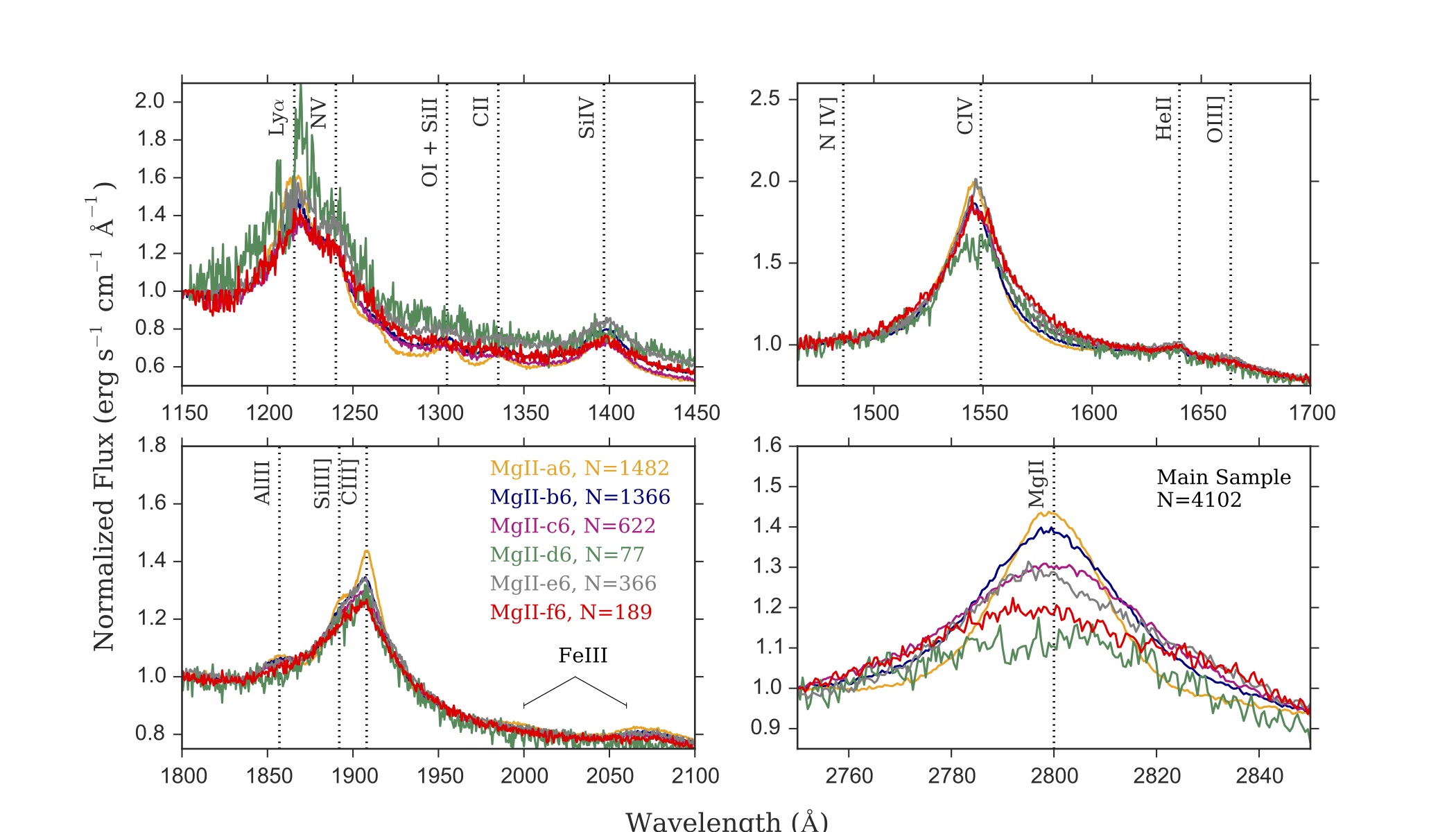}
\caption{Median composite spectra made from the objects in the \mgii\ clusters in Fig. \ref{fig:mg2_clstrs}
Similar to Fig. \ref{fig:mg2_k5_profiles} but for $K=6$.}
	\label{fig:mg2_k6_profiles}
\end{figure*}

\subsection{Main Sample -\civ}

\begin{figure*}
\includegraphics[width=\linewidth]{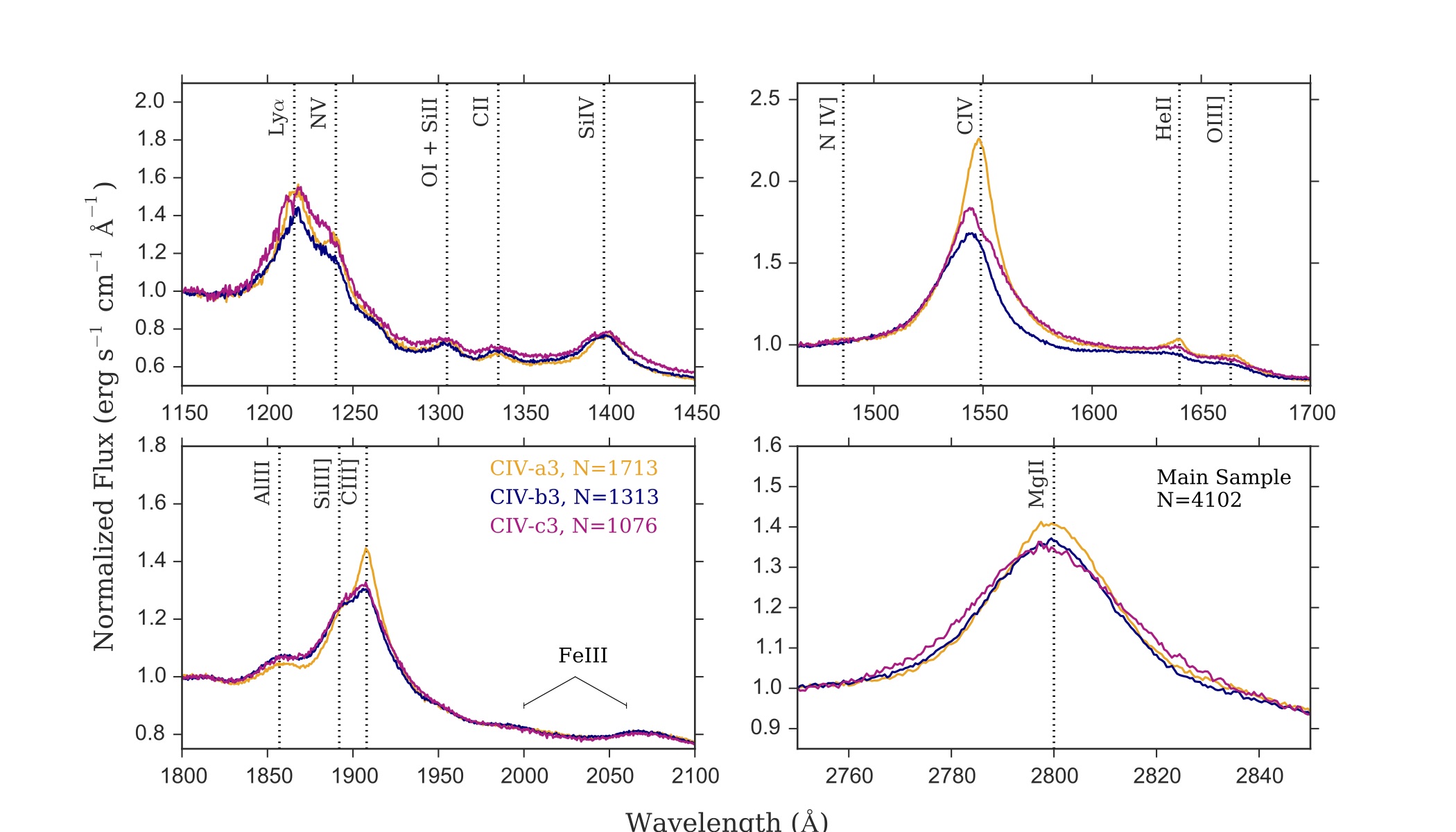}
\caption{Median composite spectra made from the objects in the \civ\ clusters.
Similar to Fig. \ref{fig:c4_k5_profiles} but for $K=3$.}
	\label{fig:c4_k3_profiles}
\end{figure*}

\begin{figure*}
\includegraphics[width=\linewidth]{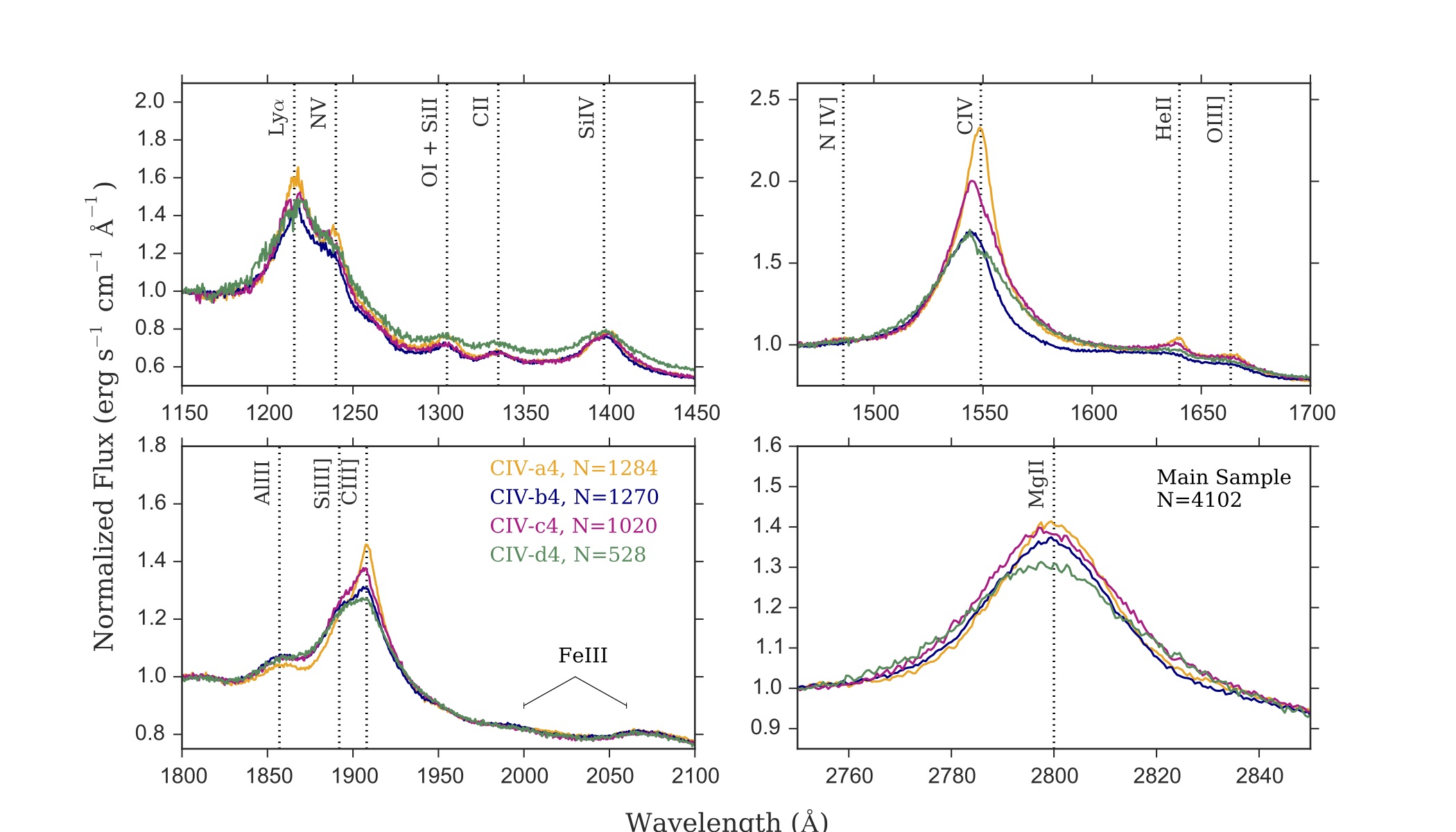}
\caption{Median composite spectra made from the objects in the \civ\ clusters.
Similar to Fig. \ref{fig:c4_k5_profiles} but for $K=4$.}
	\label{fig:c4_k4_profiles}
\end{figure*}

\begin{figure*}
\includegraphics[width=\linewidth]{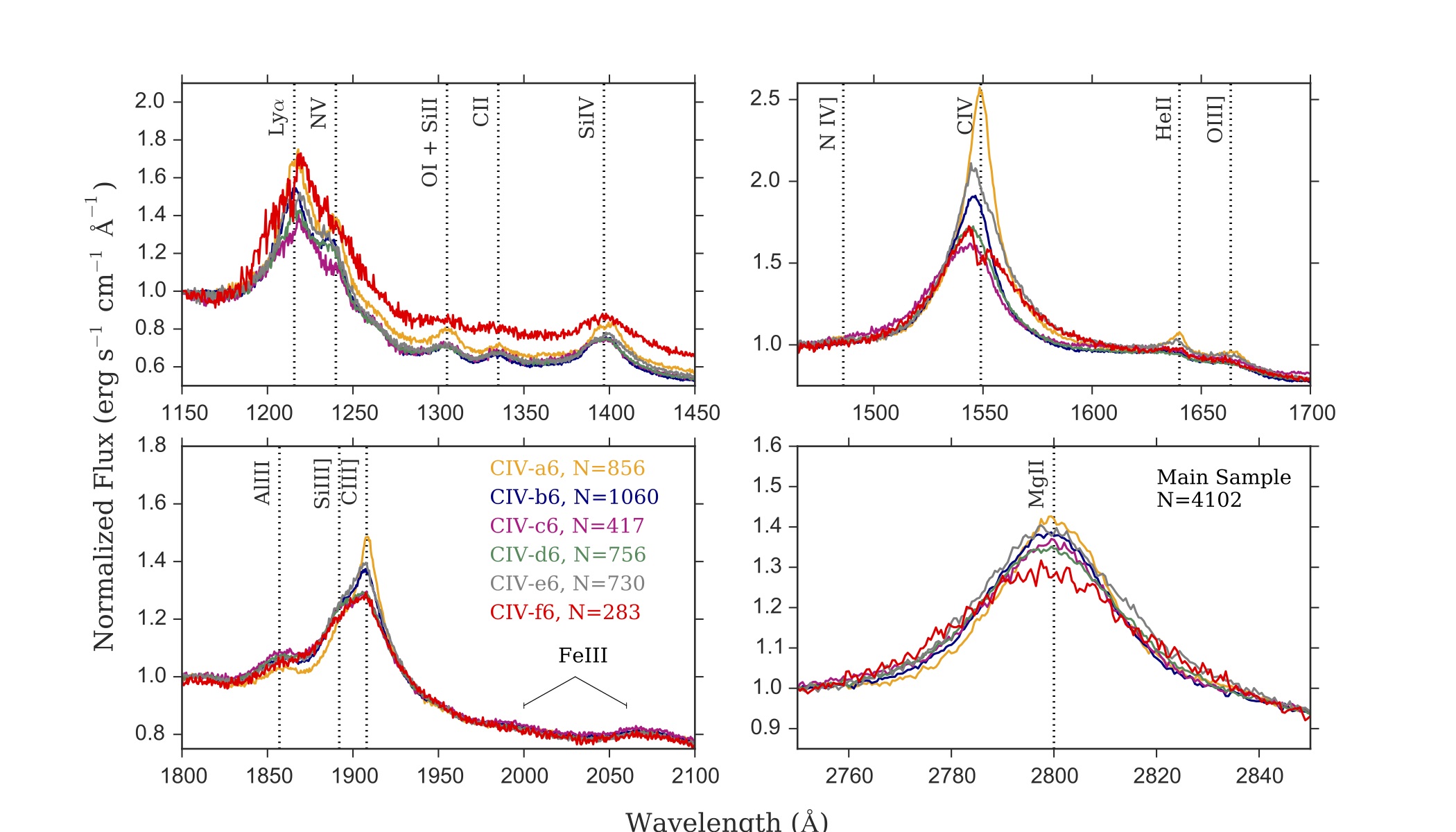}
\caption{Median composite spectra made from the objects in the \civ\ clusters.
Similar to Fig. \ref{fig:c4_k5_profiles} but for $K=6$.}
	\label{fig:c4_k6_profiles}
\end{figure*}

\subsection{Main sample -\ciii}
\begin{figure*}
\includegraphics[width=\linewidth]{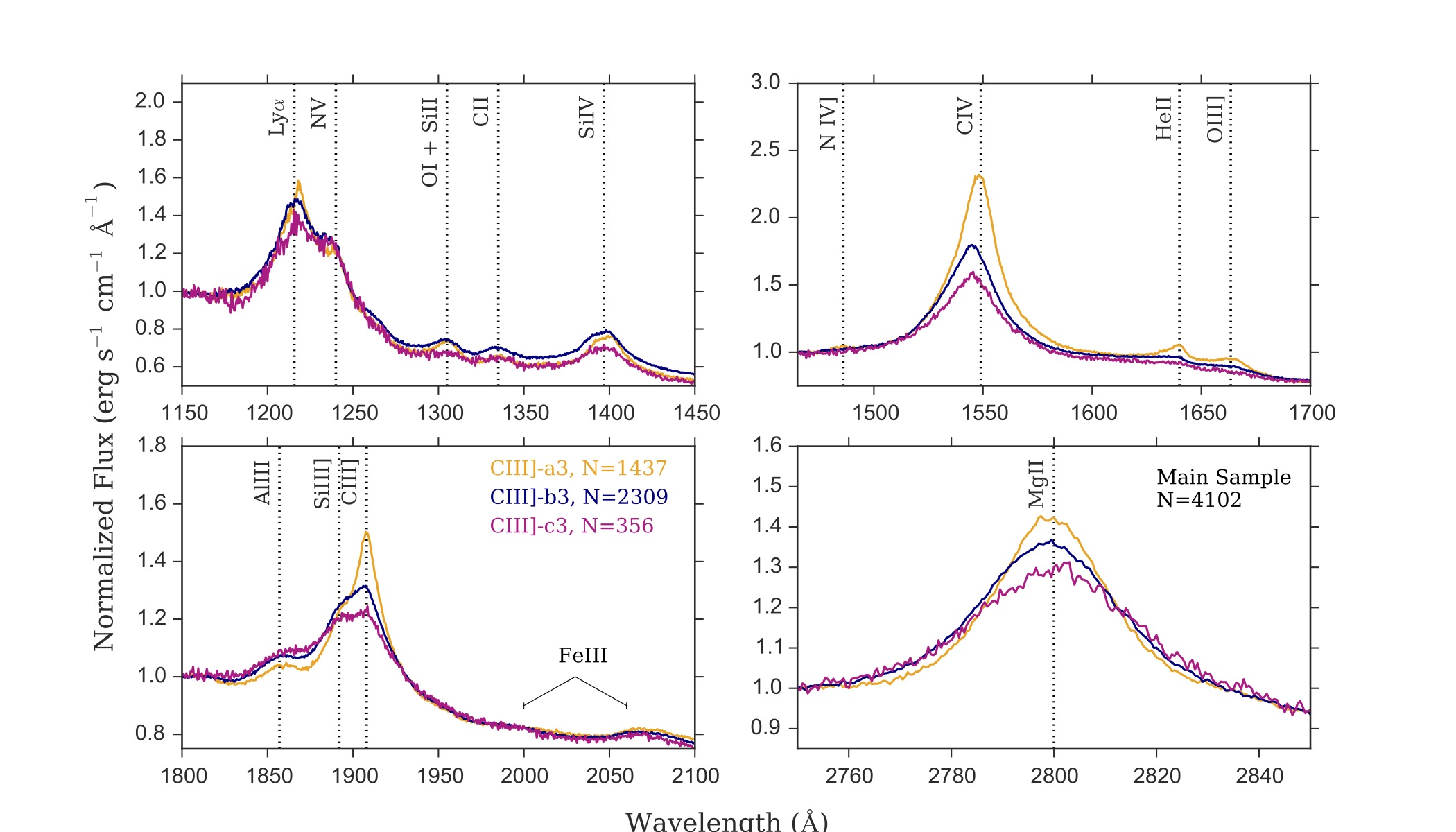}
\caption{Median composite spectra made from the objects in the \ciii\ clusters.
Similar to Fig. \ref{fig:c3_k5_profiles} but for $K=3$.}
	\label{fig:c3_k3_profiles}
\end{figure*}

\begin{figure*}
\includegraphics[width=\linewidth]{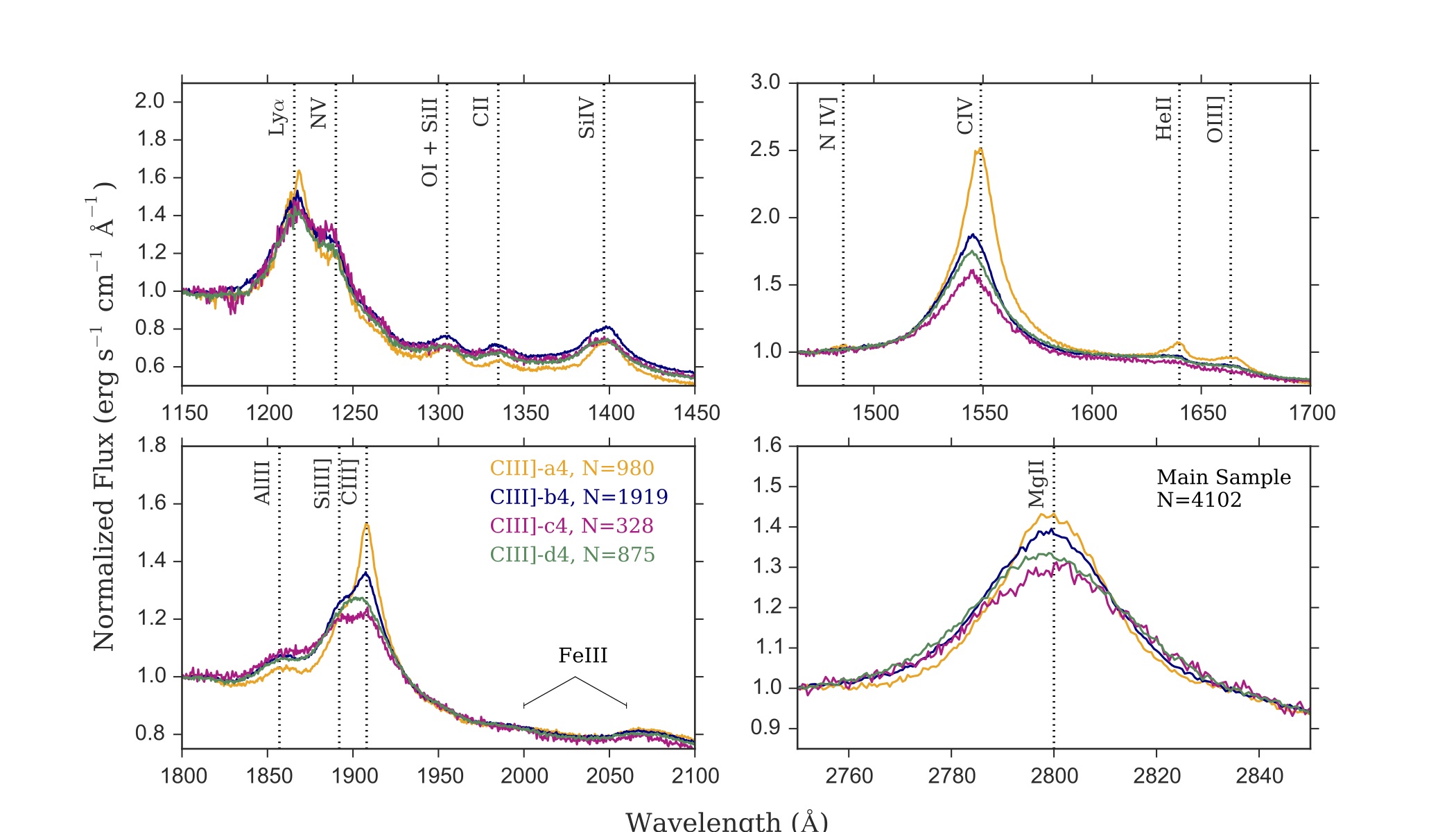}
\caption{Median composite spectra made from the objects in the \ciii\ clusters.
Similar to Fig. \ref{fig:c3_k5_profiles} but for $K=4$.}
	\label{fig:c3_k4_profiles}
\end{figure*}

\begin{figure*}
\includegraphics[width=\linewidth]{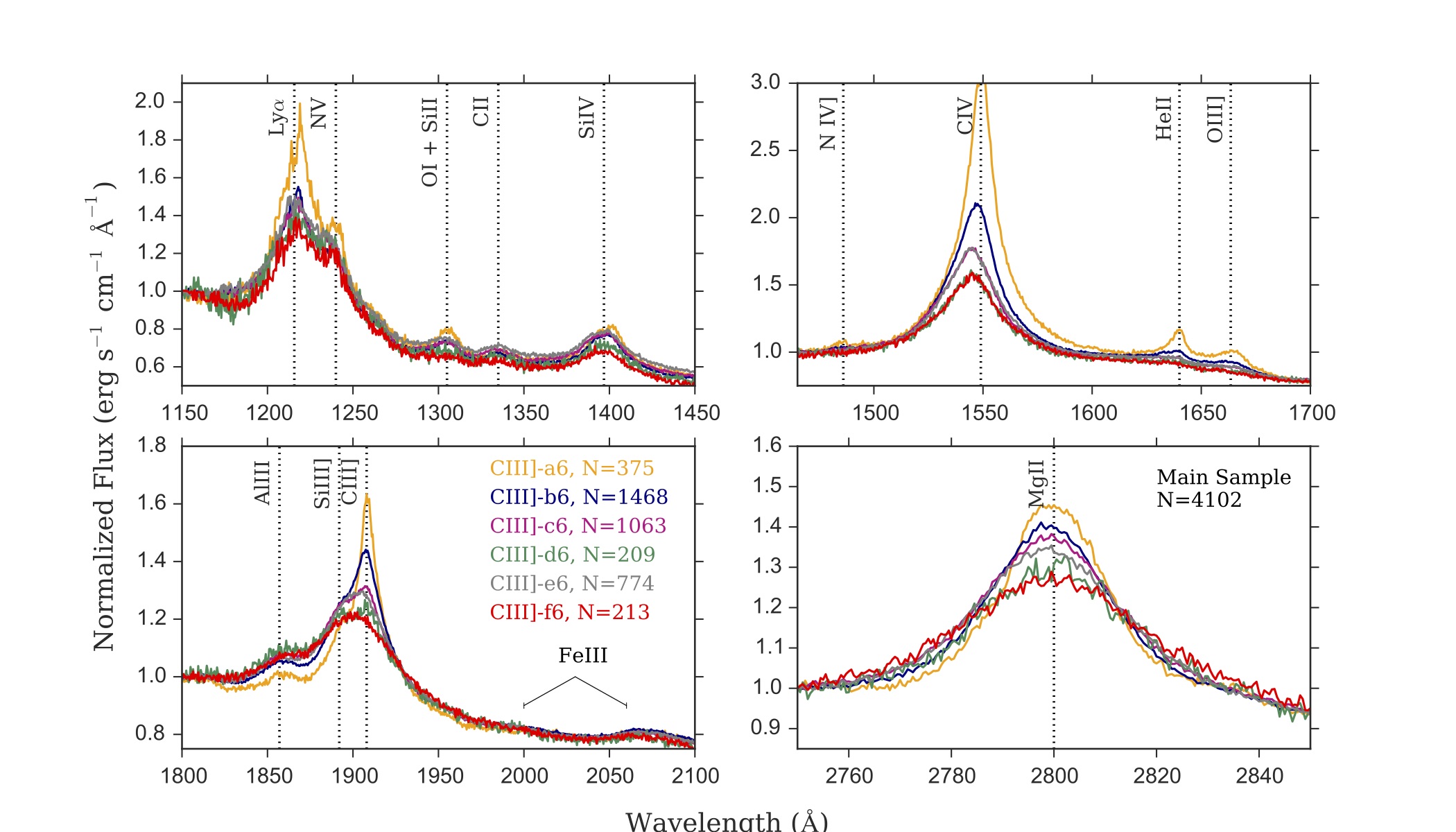}
\caption{Median composite spectra made from the objects in the \ciii\ clusters.
Similar to Fig. \ref{fig:c3_k5_profiles} but for $K=6$.}
	\label{fig:c3_k6_profiles}
\end{figure*}

\subsection{Mixed Sample}

\begin{figure*}
\includegraphics[width=\linewidth]{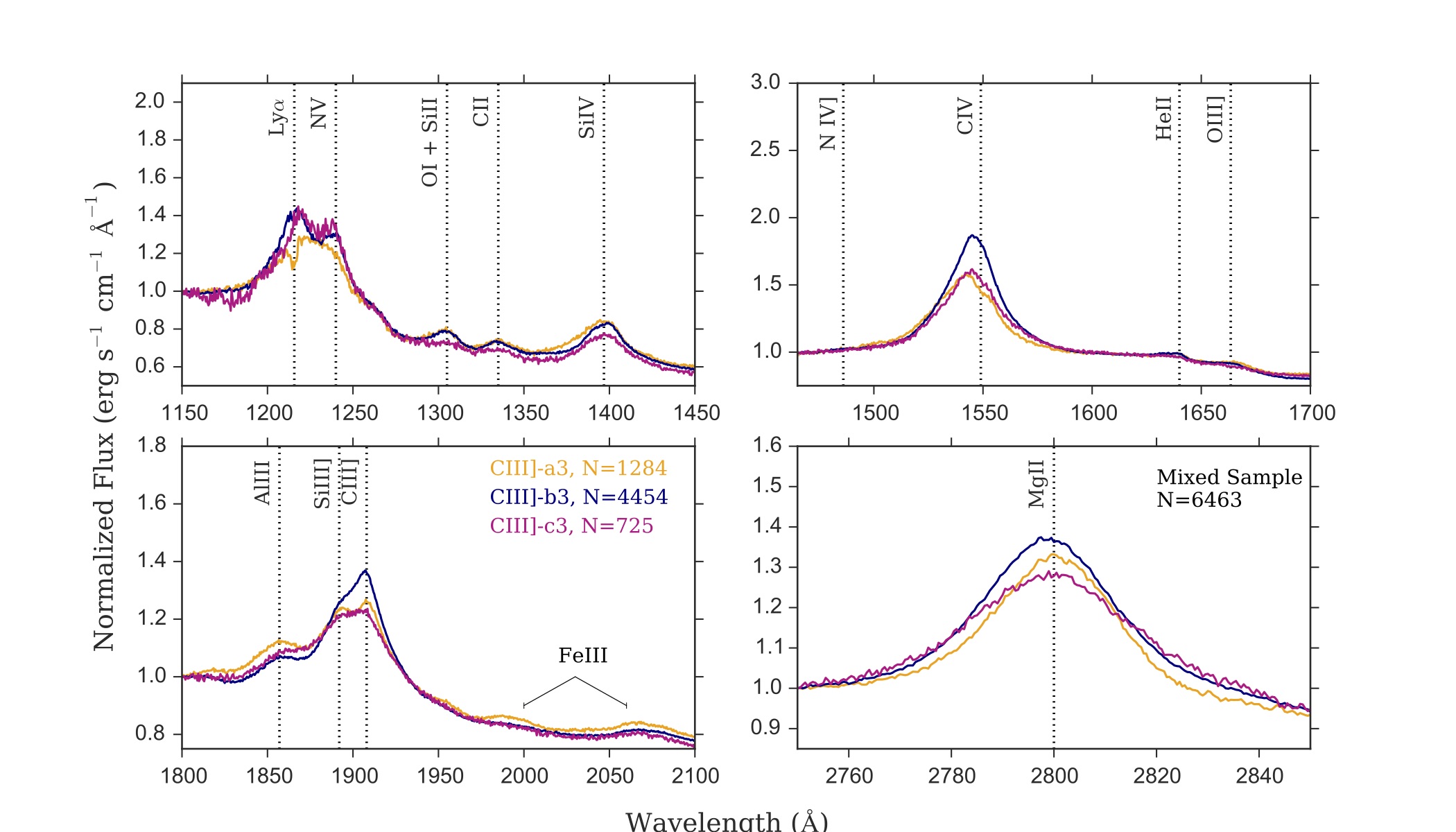}
\caption{Median composite spectra made from the objects in the \ciii\ clusters in the mixed sample. 
Similar to Fig. \ref{fig:c3_k6_mixed_profiles} but for $K=3$.}
	\label{fig:c3_k3_mixed_profiles}
\end{figure*}

\begin{figure*}
\includegraphics[width=\linewidth]{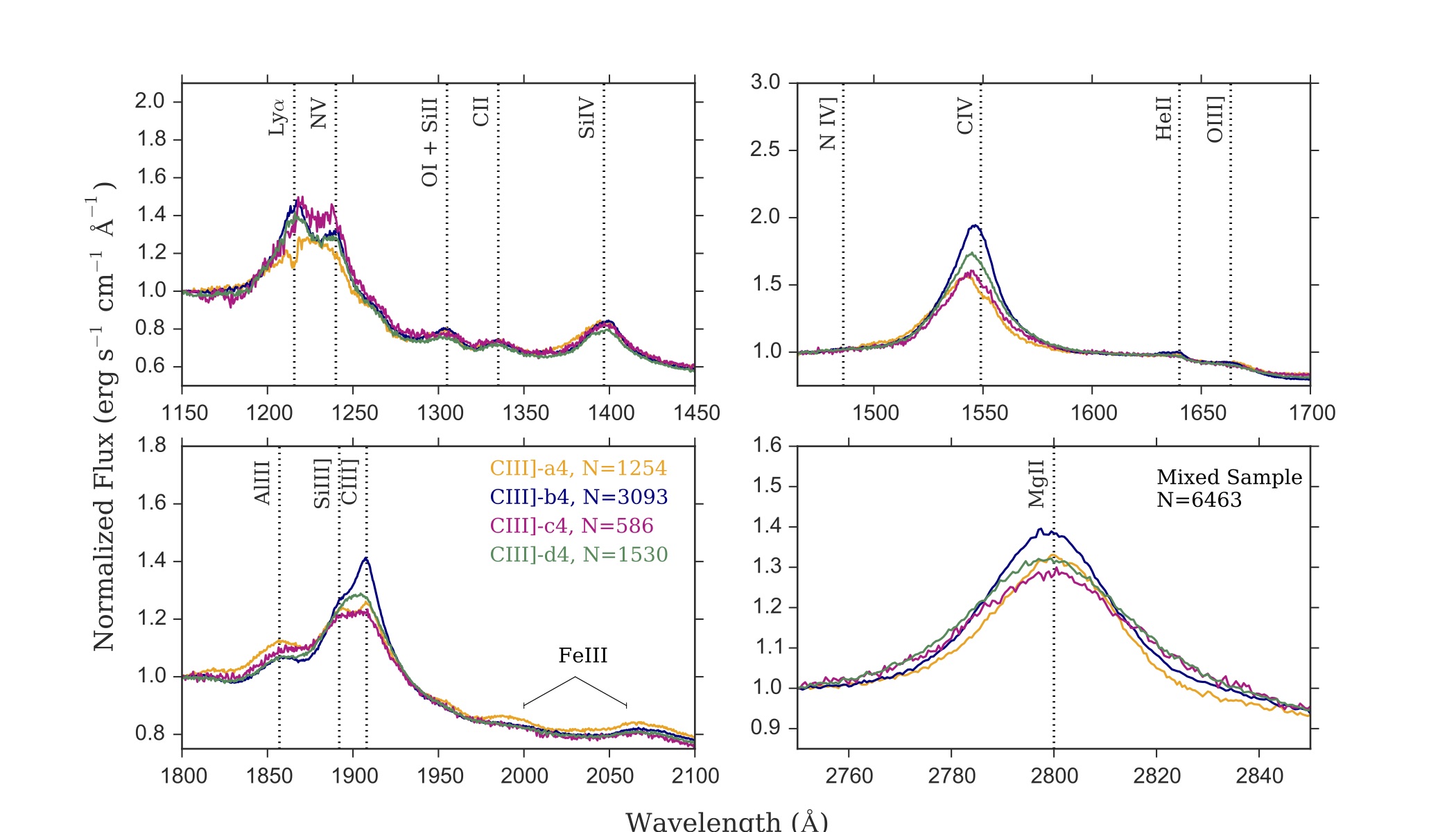}
\caption{Median composite spectra made from the objects in the \ciii\ clusters in the mixed sample. 
Similar to Fig. \ref{fig:c3_k6_mixed_profiles} but for $K=4$.}
	\label{fig:c3_k4_mixed_profiles}
\end{figure*}

\begin{figure*}
\includegraphics[width=\linewidth]{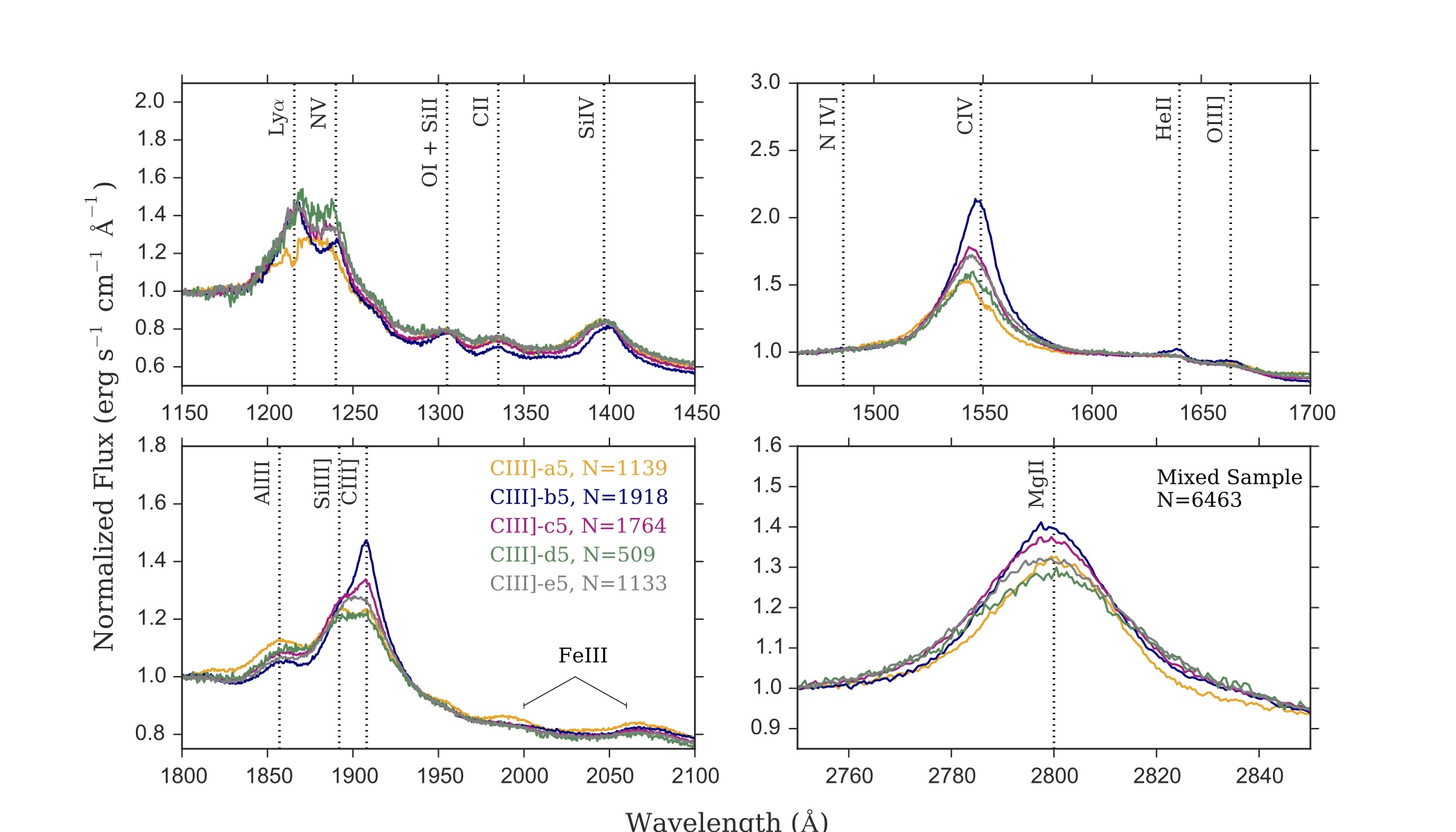}
\caption{Median composite spectra made from the objects in the \ciii\ clusters in the mixed sample. 
Similar to Fig. \ref{fig:c3_k6_mixed_profiles} but for $K=5$.}
	\label{fig:c3_k5_mixed_profiles}
\end{figure*}

\subsection{BAL Quasars Only}
\begin{figure*}
\includegraphics[width=\linewidth]{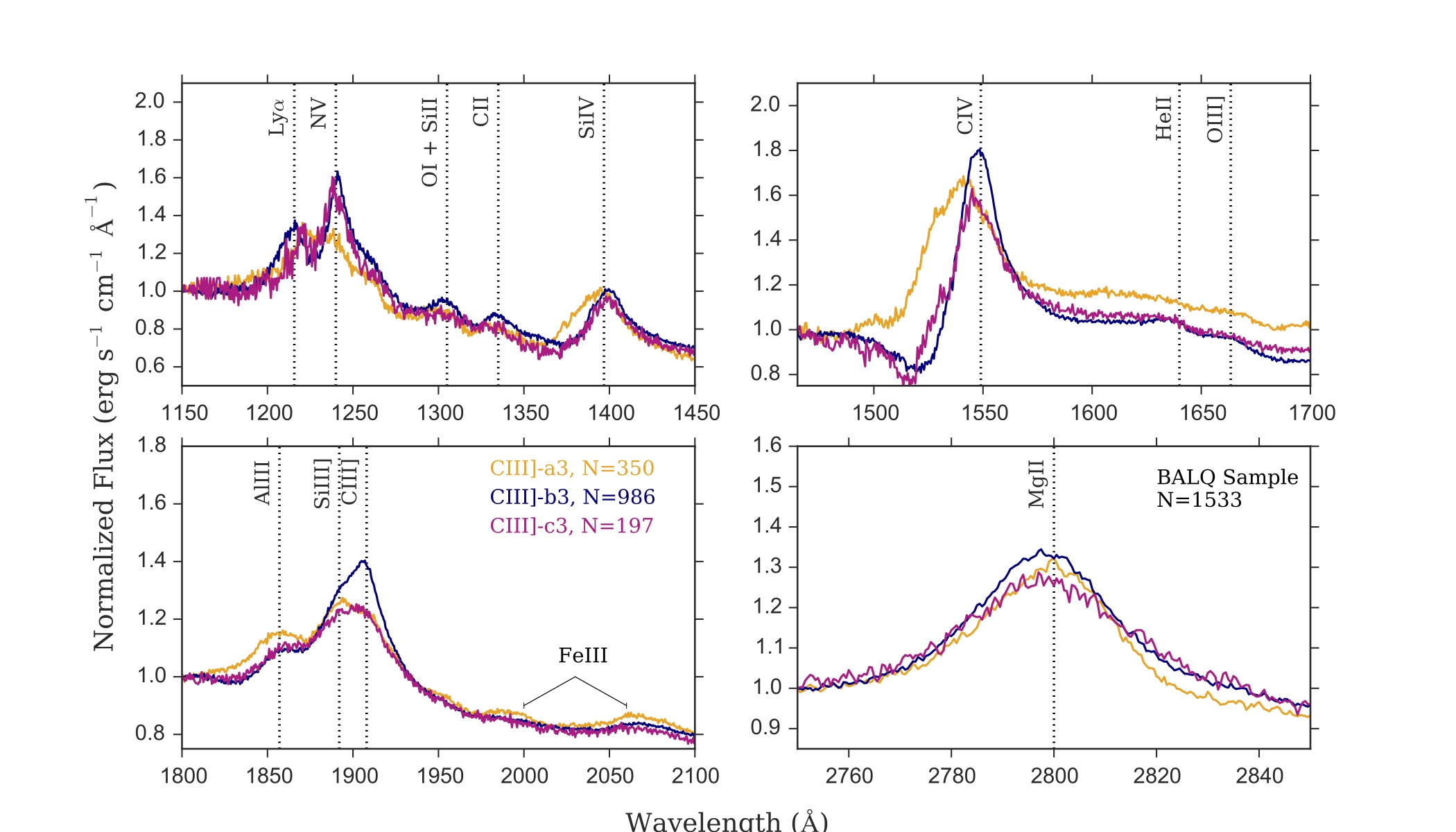}
\caption{Median composite spectra made from the objects in the \ciii\ clusters in the BALQs sample. 
Similar to Fig. \ref{fig:c3_k5_bal_profiles} but for $K=3$.}
	\label{fig:c3_k3_bal_profiles}
\end{figure*}

\begin{figure*}
\includegraphics[width=\linewidth]{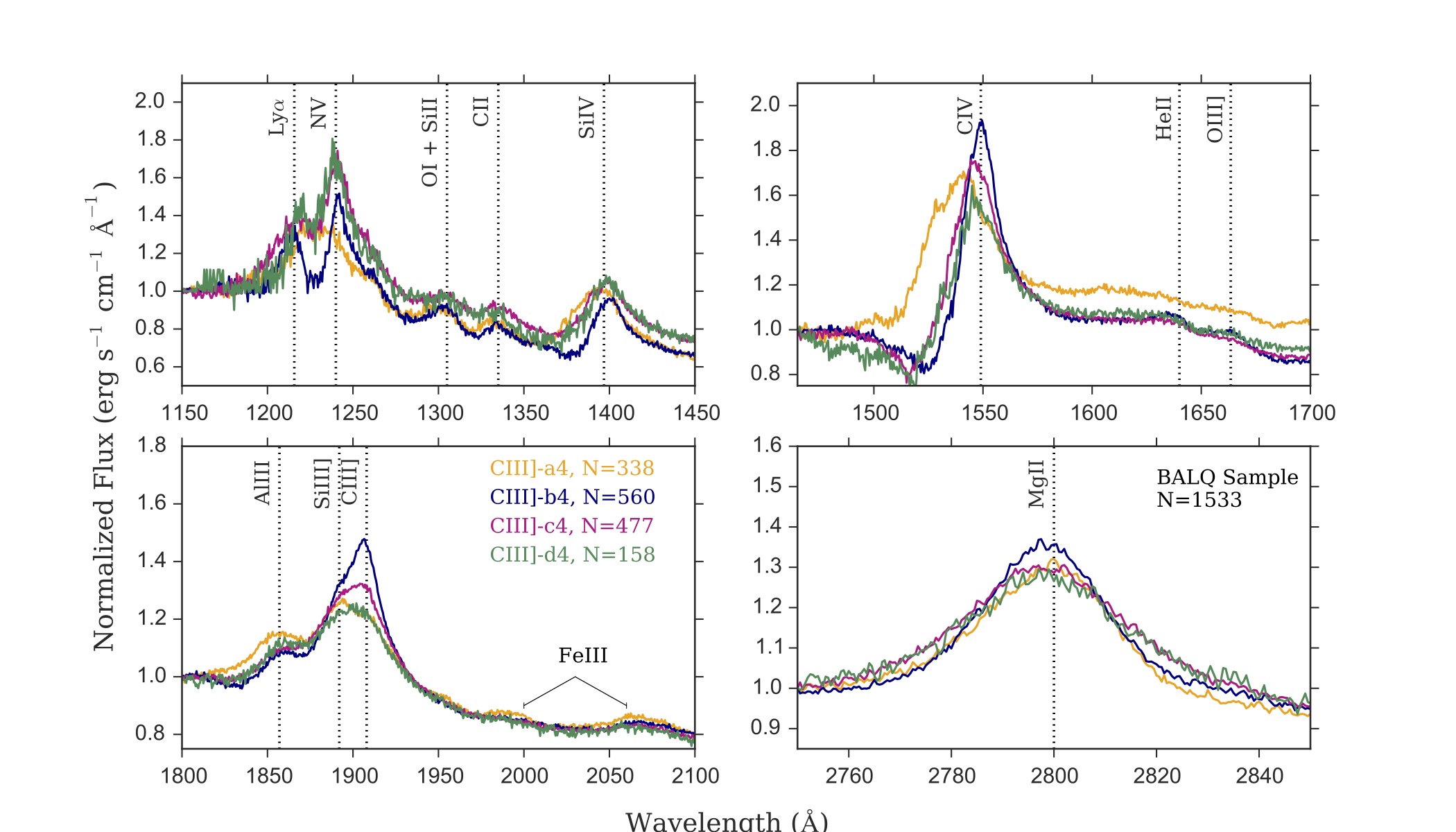}
\caption{Median composite spectra made from the objects in the \ciii\ clusters in the BALQs sample. 
Similar to Fig. \ref{fig:c3_k5_bal_profiles} but for $K=4$.}
	\label{fig:c3_k4_bal_profiles}
\end{figure*}

\begin{figure*}
\includegraphics[width=\linewidth]{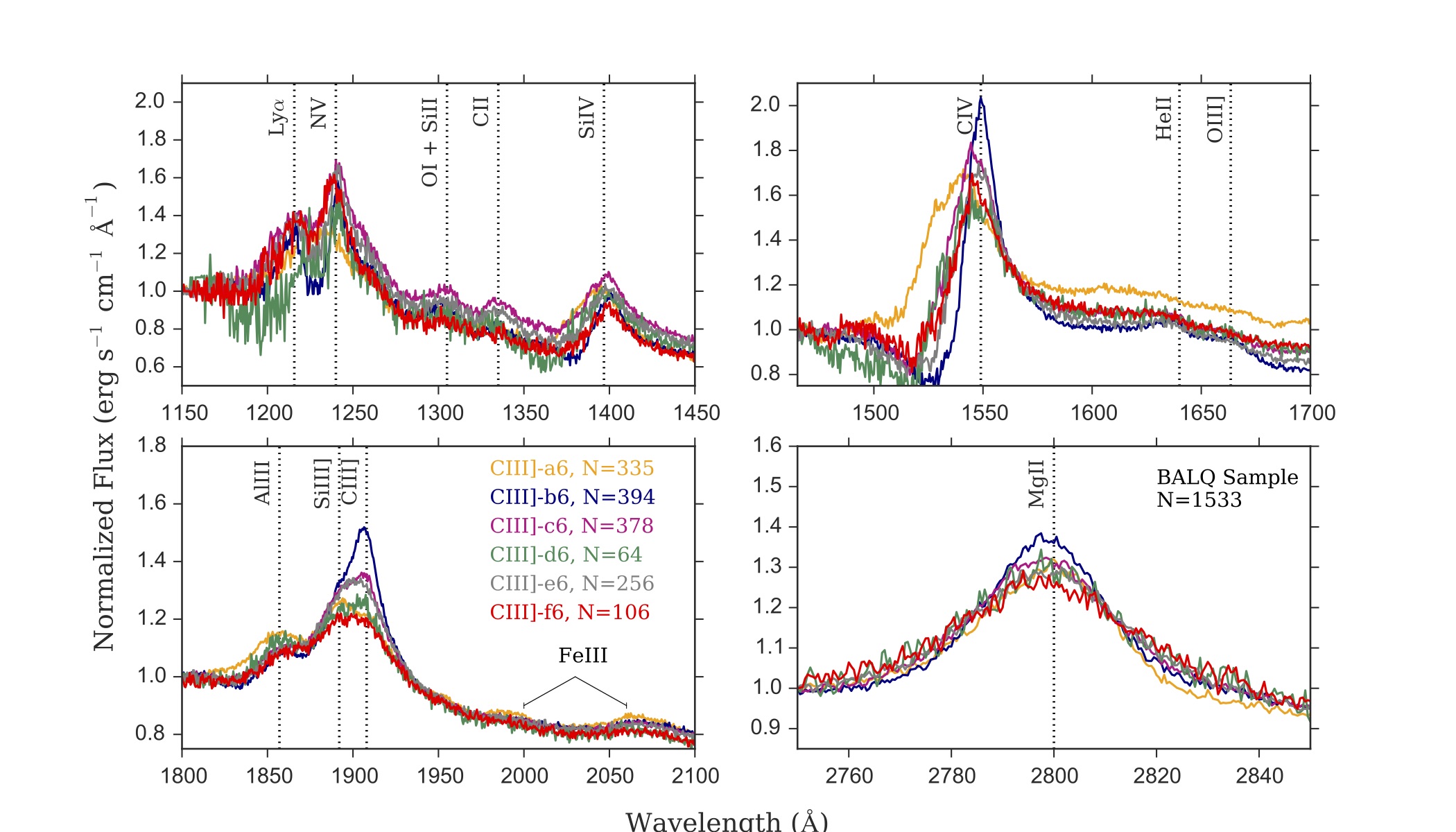}
\caption{Median composite spectra made from the objects in the \ciii\ clusters in the BALQs sample. 
Similar to Fig. \ref{fig:c3_k5_bal_profiles} but for $K=6$.}
	\label{fig:c3_k6_bal_profiles}
\end{figure*}

\clearpage
\section{Reproducibility of Clusters}
\label{sec:rep_plots}
We tested whether the algorithm is able to find the same unique set of centroids when the clustering is repeated.
We find that for the clusters generated in the main sample, the algorithm is finding the same set of clusters even after 50 repeats.
 We show here one example for $K = 5$.

\begin{figure*}
  \includegraphics[width=\linewidth]{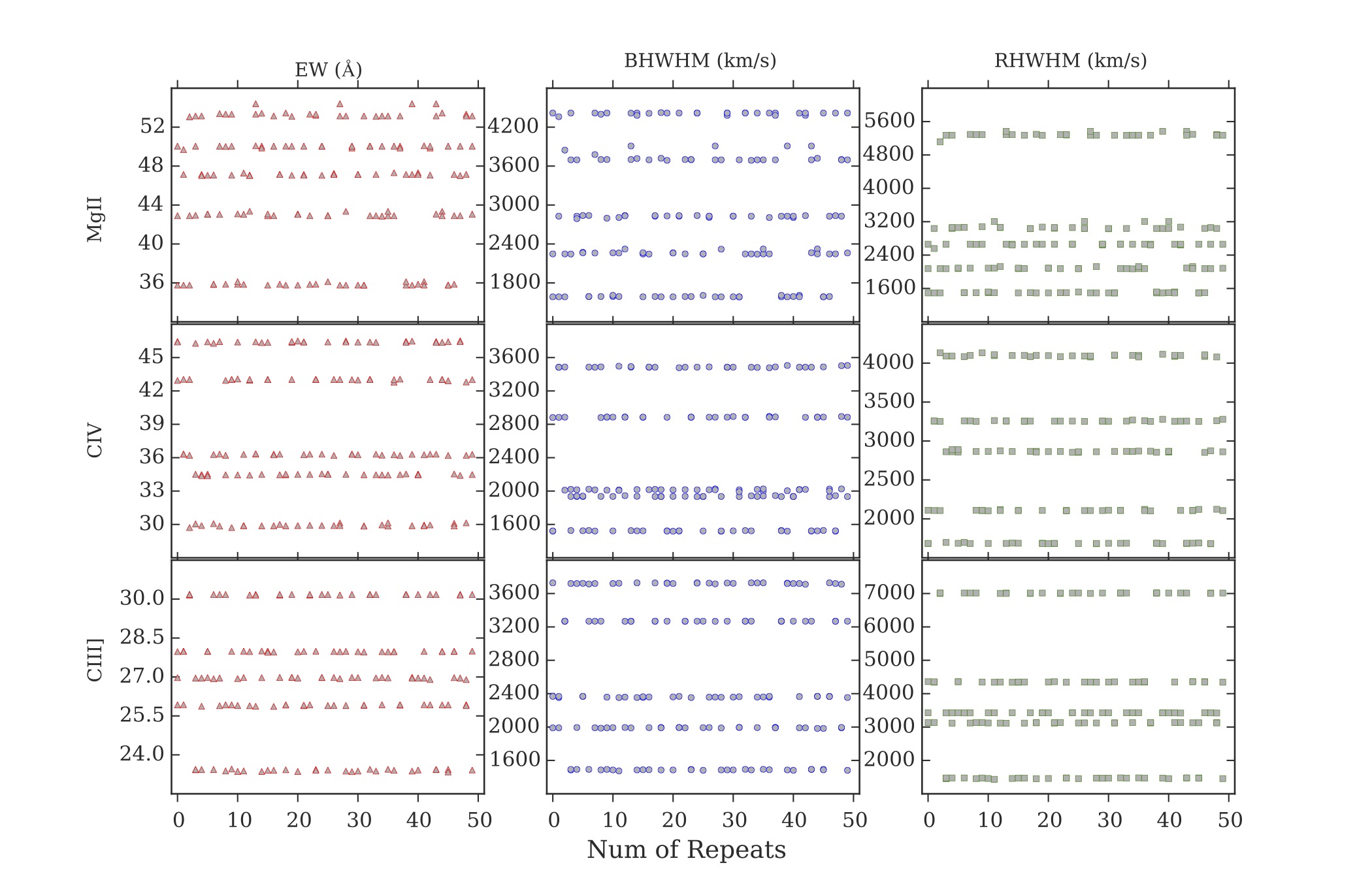}
  \caption{The centroids for three different clustering attempts one for each of the lines (\mgii, \civ, and \ciii) in the main sample repeated 50 times using $K=5$.
  This shows that K-means is finding clusters that are consistent and that the clusters we use in our analysis reflect meaningful groupings of like objects in the parameter space of EW, RHWHM and BHWHM.}
 \label{fig:cntrs_k5}
\end{figure*}

\newpage

\bibliographystyle{mnras}
\bibliography{qclusters}

\end{document}